\documentclass[preprint]{aastex}
\slugcomment{Published in PASP, 118, 62 (Jan. 2006)}
\shorttitle{Theoretical Isochrones near $K$ Band}
\shortauthors{KIM, FIGER, \& LEE}

\def\spose#1{\hbox to 0pt{#1\hss}}
\newcommand\lsim{\mathrel{\spose{\lower 3.0pt\hbox{$\mathchar"218$}}
     \raise 2.0pt\hbox{$\mathchar"13C$}}}
\newcommand\gsim{\mathrel{\spose{\lower 3.0pt\hbox{$\mathchar"218$}}
     \raise 2.0pt\hbox{$\mathchar"13E$}}}

\begin{document}
\title{Theoretical Isochrones with Extinction in the $K$ Band. II. \\
$J$ -- $K$ versus $K$}
\author{Sungsoo S. Kim\altaffilmark{1}, Donald F. Figer\altaffilmark{2}, and
Myung Gyoon Lee\altaffilmark{3}}
\altaffiltext{1}{Department of Astronomy and Space Science, Kyung Hee
University, Yongin-shi, Kyungki-do 449-701, South Korea; sungsoo.kim@khu.ac.kr.}
\altaffiltext{2}{Space Telescope Science Institute, 3700 San Martin Drive,
Baltimore, MD 21218; figer@stsci.edu.}
\altaffiltext{3}{Astronomy Program, SEES, Seoul National University,
Seoul 151-742, South Korea; mglee@astrog.snu.ac.kr.}

\begin{abstract}
We calculate theoretical isochrones in a consistent way for five
filter pairs near the $J$ and $K$ band atmospheric windows 
($J$--$K$, $J$--$K'$,
$J$--$K_s$, F110W--F205W, and F110W--F222M) using the Padova stellar
evolutionary models of Girardi et al.  We present magnitude transformations
between various $K$-band filters as a function of color. 
Isochrones with extinction of up to 6 mag in the $K$ band are also
presented.  As found for the filter pairs composed of $H$ \& $K$ band filters,
we find that the reddened isochrones of different filter pairs behave
as if they follow different extinction laws, and that the extinction
curves of {\it Hubble Space Telescope} NICMOS filter pairs in the
color-magnitude diagram are
considerably nonlinear.  Because of these problems, extinction
values estimated with NICMOS filters can be in error by up to 1.3 mag.
Our calculation suggests that the extinction
law implied by the observations of Rieke et al for wavelengths
between the $J$ and $K$ bands is better described by a power-law function
with an exponent of 1.66 instead of 1.59, which is commonly used with
an assumption that the transmission functions of $J$ and $K$ filters
are Dirac delta functions.
\end{abstract}

\section{INTRODUCTION}
\label{sec:introduction}

Kim et al. (2005, hereafter Paper I) have calculated theoretical isochrones
with extinction for some $H$ and $K$ band filters using the Padova stellar
evolutionary models by Girardi et al. (2002).  In Paper I, we found that the
reddened isochrones of different filter pairs in $H$ and $K$ bands behave
as if they follow different extinction laws, and that care is needed when
applying an extinction law obtained with one filter pair to other,
similar filter pairs.  For example, if the extinction law for
the Johnson-Glass $H$ and $K$ filters obtained by Rieke, Rieke, \& Paul (1989)
is directly applied to the photometry from the {\it Hubble Space
Telescope} ({\it HST}) NICMOS filters
(F160W, F205W, and F222M), estimated extinction values can be in error
by up to 0.3 mag for true extinction at $K$ of 6 mag or less.  To reduce
this error, Paper I introduced an ``effective extinction slope'' for each
filter pair and isochrone model.  It was also found that the extinction
behavior of isochrones in the color-magnitude diagram (CMD) for filter pair
F160W--F222M is highly nonlinear (i.e., the amount of extinction is not
proportional to color excess) because of a significant width difference
in the two filters.

\begin{figure}
\epsscale{0.9}
\plotone{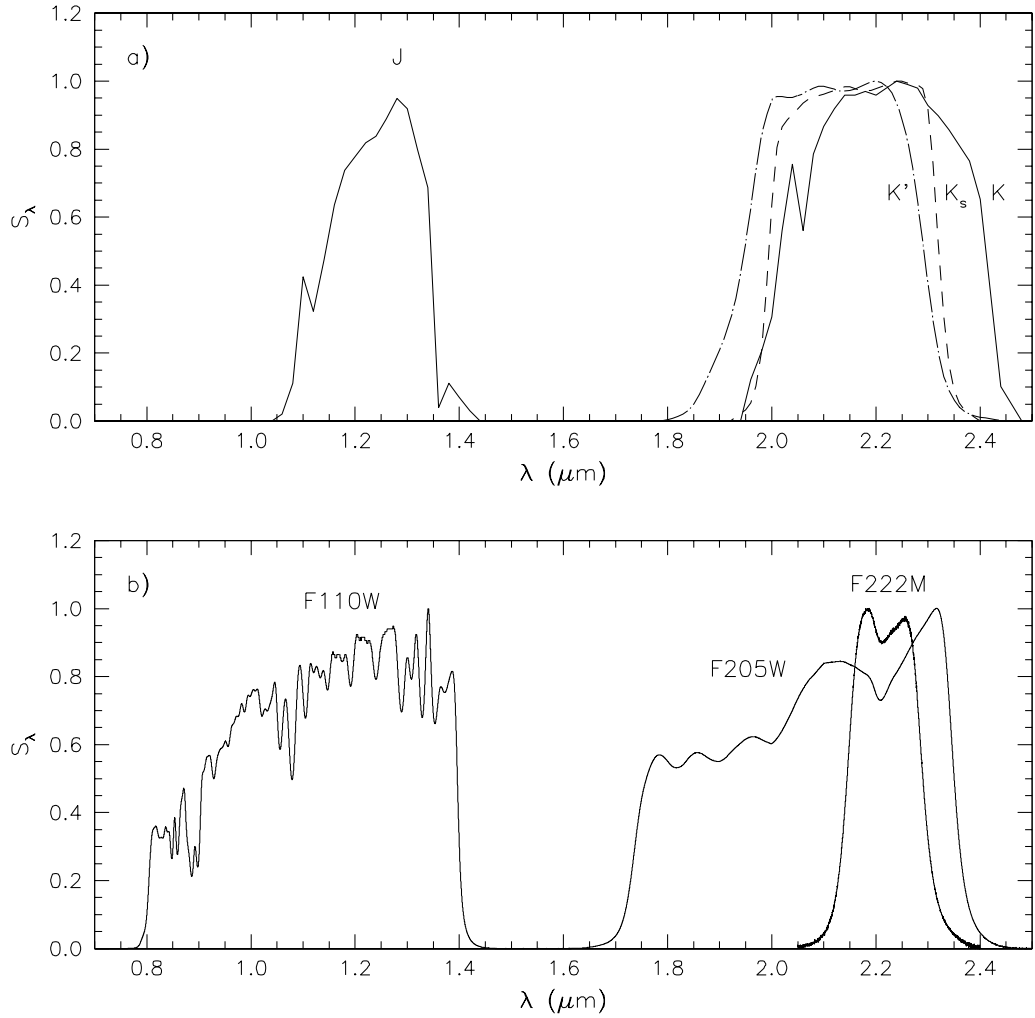}
\caption
{\label{fig:filter}Transmission functions ($S_\lambda$) of the filters
considered in the present work.  The $S_\lambda$ values are scaled such
that their maximums occur at 1.}
\end{figure}

These problems are certainly not limited to the isochrones for filter pairs
in the $H$ and $K$ bands.  This problem will apply to any situation in
which one applies an extinction law deduced from one filter pair to other
similar filter pairs.  Furthermore, the nonlinear behavior of the extinction
vector in the CMD will be problematic for filter pairs
with significant difference in width.  In the present paper, we extend
the calculations performed in Paper I to the isochrones for filter pairs
in the $J$ and $K$ bands.  The filters considered here are the four
ground-based filters $J$, $K$ (Johnson et al. 1966), $K'$ (Wainscoat
\& Cowie 1992), and $K_s$ ($K$-short; developed by M. Skrutskie; see the
appendix of Persson et al. 1998), and the three NICMOS filters
F110W, F205W, and F222M (transmission functions of these filters are
shown in Figure~\ref{fig:filter}).  Out of these seven filters, we consider
five filter pairs: $J$--$K$, $J$--$K'$, $J$--$K_s$, F110W--F205W, and
F110W--F222M.

We adopt a Vega-based photometric system (VEGAMAG system), which uses
Vega ($\alpha$ Lyr) as the calibrating star.  For photometric zero
points of NICMOS filters, we adopt $\langle f_\nu^{\rm Vega} \rangle$
values from the NICMOS Data Handbook (ver. 5.0): 1775~Jy for F110W,
703.6~Jy for F205W, and 610.4~Jy for F222M.  For the spectra of
synthetic stellar atmospheres, we adopt Kurucz ATLAS9 no-overshoot
models\footnote{See NOVER files at http://kurucz.harvard.edu/grids.html.}
(Kurucz 1993) calculated by Castelli et al. (1997).  The metallicities of
these models cover the values of [M/H] = $-2.5$ to $+0.5$.  A microturbulent
velocity $\xi=2\,{\rm km \, s^{-1}}$ and a mixing length parameter
$\alpha =1.25$ are adopted in the present study.  For the temporal
evolution of effective temperature and luminosity as functions of stellar
mass (i.e., stellar evolutionary tracks), we adopt the ``basic set" of
the Padova models\footnote{See http://pleiadi.pd.astro.it.} (Girardi et al.
2002).  We consider isochrones with a metallicity $Z$ = 0.0001, 0.001, 0.019,
and 0.03.  The stellar spectral library and the evolutionary tracks we
adopted assume a solar chemical ratios.

For more details on the magnitude system, stellar spectral library, and
evolutionary tracks that we adopt here, readers are referred to Paper I.

Throughout this paper, we generically refer to the atmospheric wavebands
centered near 1.25, 1.65, and 2.2~\micron, as the $J$, $H$,
and $K$ bands, whereas we refer to the Johnson-Glass filters (Johnson et al.
1966; Glass 1974) as the $J$, $H$, and $K$ filters.

\section{Isochrones}
\label{sec:isochrones}

We first prepare a table of magnitudes for all spectra in ATLAS9 models in
the $J$ and $K$ band filters, covering a large range in $T_{eff}$,
$\log g$, and [M/H], using equations (5) and (6) of Paper I.
We use this table as a set of interpolates for the $T_{eff}$, $\log g$, and
$Z$ values predicted by the stellar evolution models for a given age in order
to estimate synthetic isochrones.

\begin{figure}
\epsscale{1.0}
\plotone{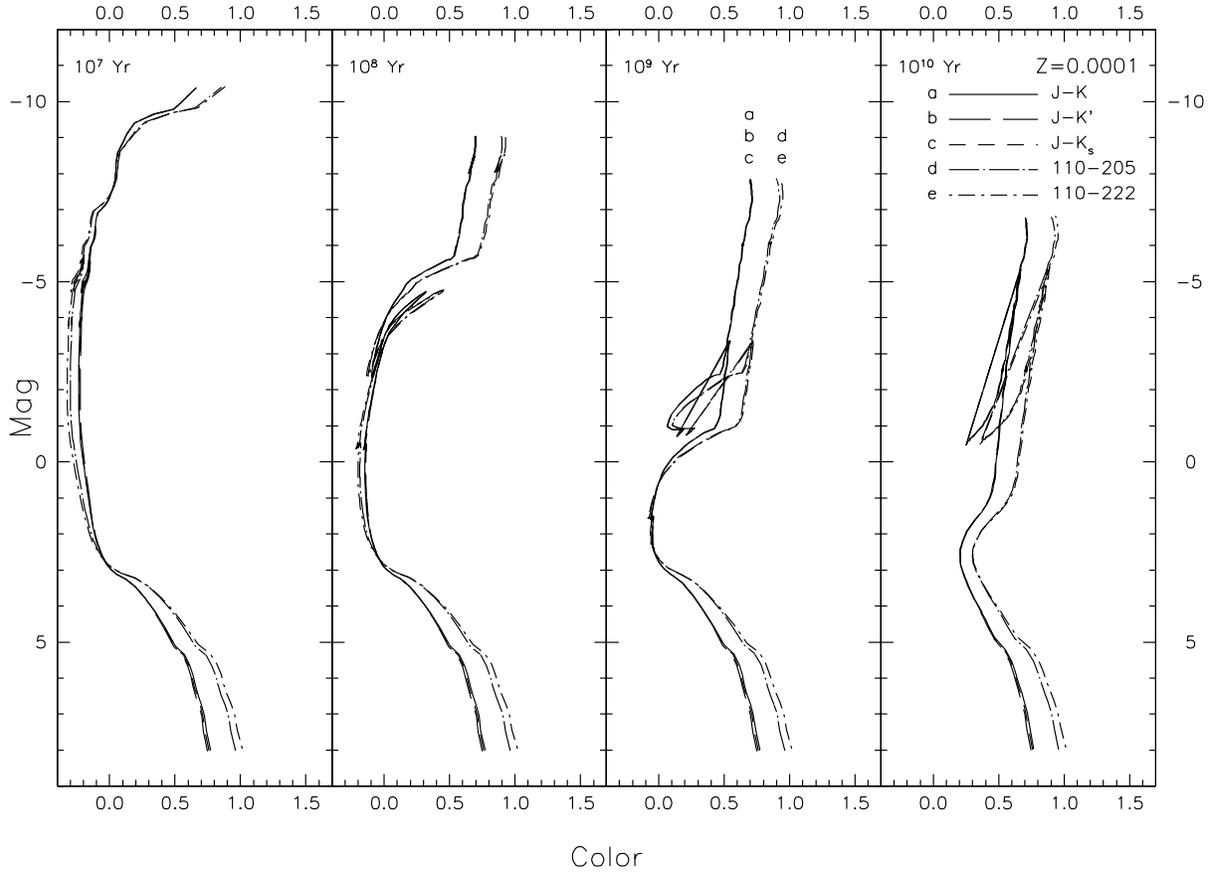}
\caption
{\label{fig:iso1}Isochrones of the $Z=0.0001$ model for $J-K$ vs. $K$ ($\it
solid line$), $J-K'$ vs. $K'$ ($\it long-dashed line$), $J-K_s$ vs. $K_s$
($\it short-dashed line$), F110W$-$F205W vs. F205W ($\it long-dash-dotted
line$), and F110W$-$F222M vs. F222M ($\it short-dash-dotted line$)
in the Vega magnitude system.  The three atmospheric filters nearly coincide
at the bright end.  Only the data points
that have $\log T_{eff} \ge 3500$~K and $\log g \ge 0$ are plotted.}
\end{figure}

\begin{figure}
\epsscale{1.0}
\plotone{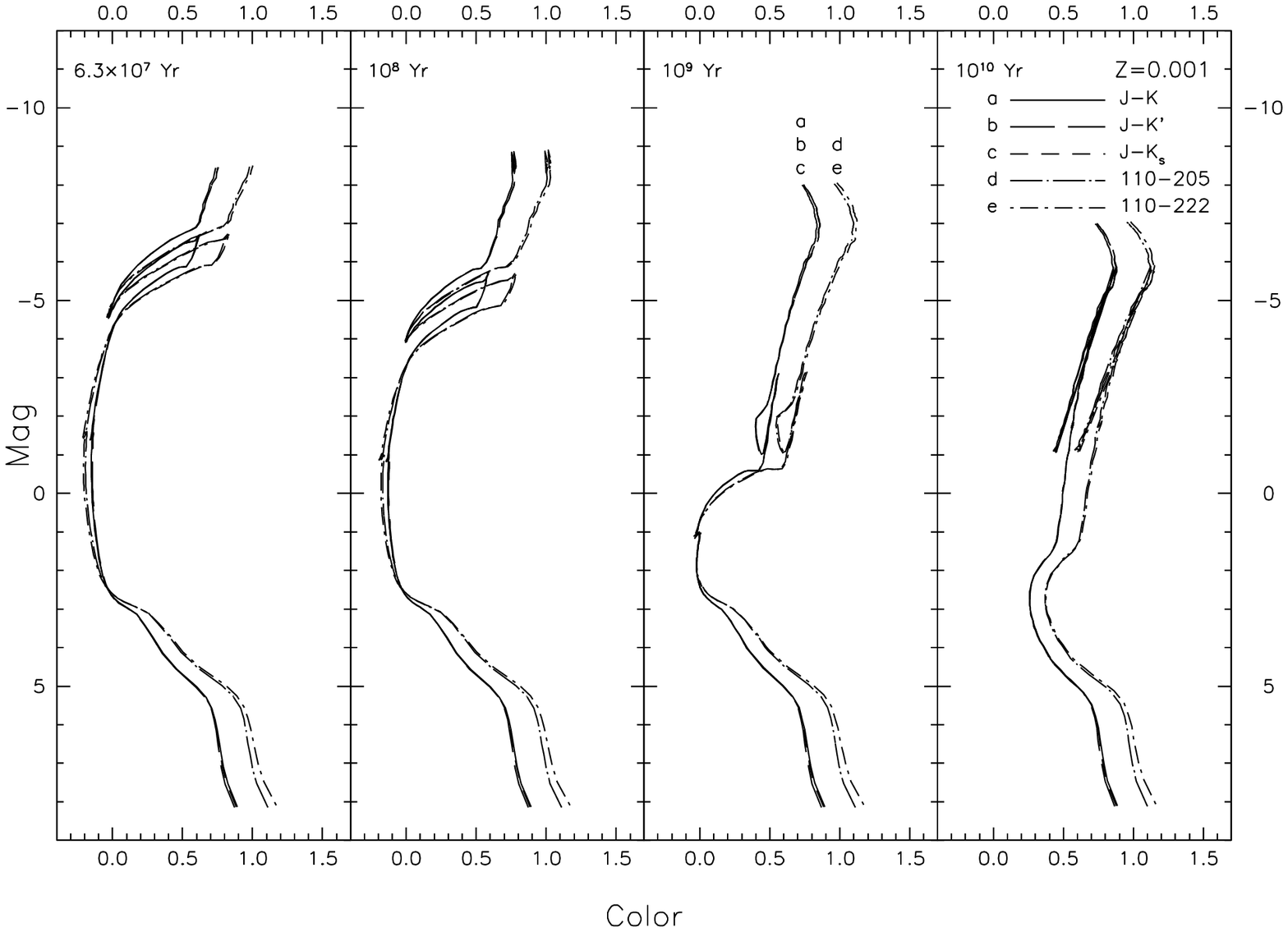}
\caption
{\label{fig:iso2}Same as Figure~\ref{fig:iso1}, but for the $Z=0.001$ model.
Isochrones for $K'$ and $K_s$ are indistinguishable.}
\end{figure}

\begin{figure}
\epsscale{1.0}
\plotone{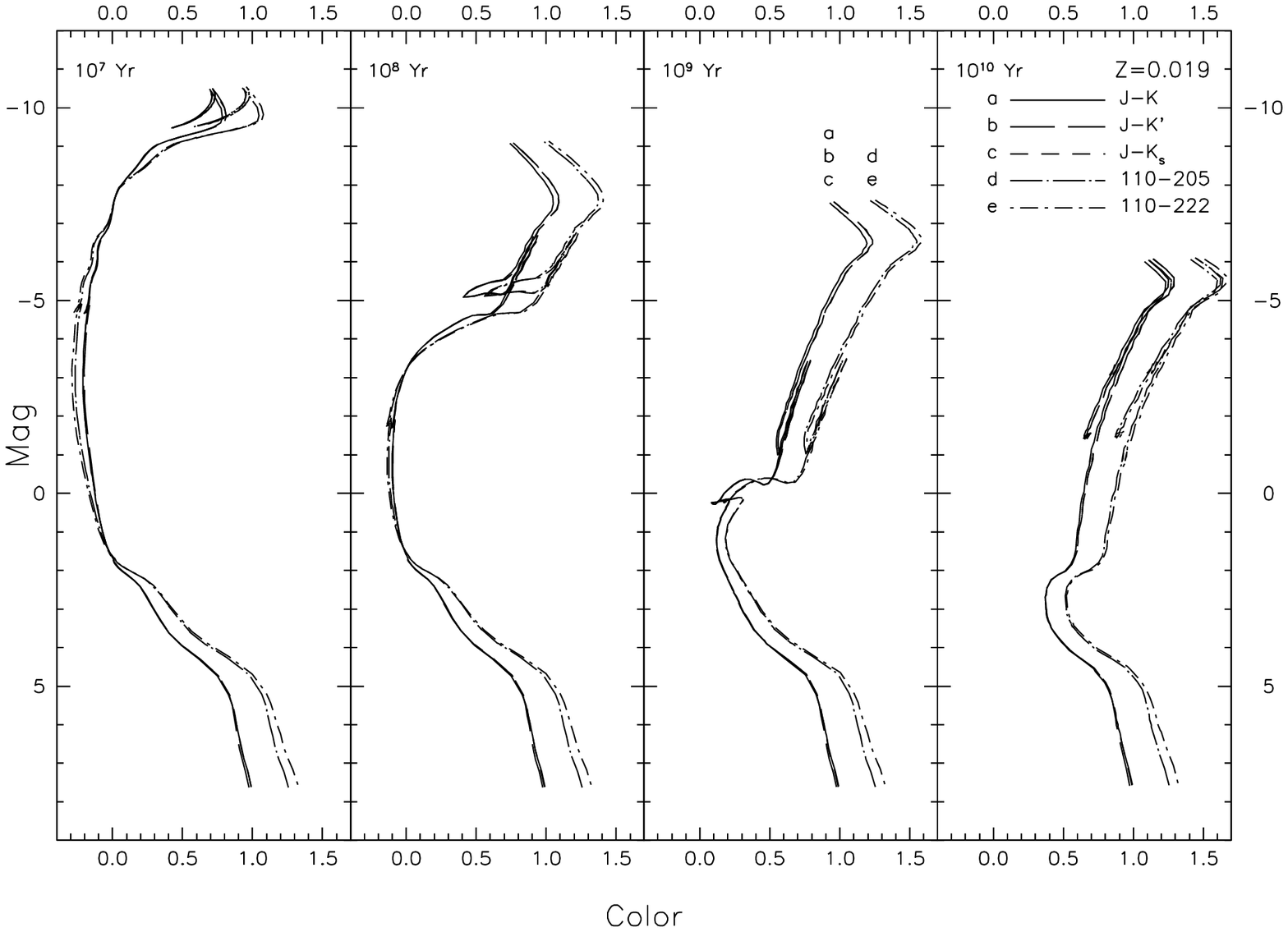}
\caption
{\label{fig:iso3}Same as Figure~\ref{fig:iso1}, but for the $Z=0.019$ model.
Isochrones for $K'$ and $K_s$ are indistinguishable.}
\end{figure}

\begin{figure}
\epsscale{1.0}
\plotone{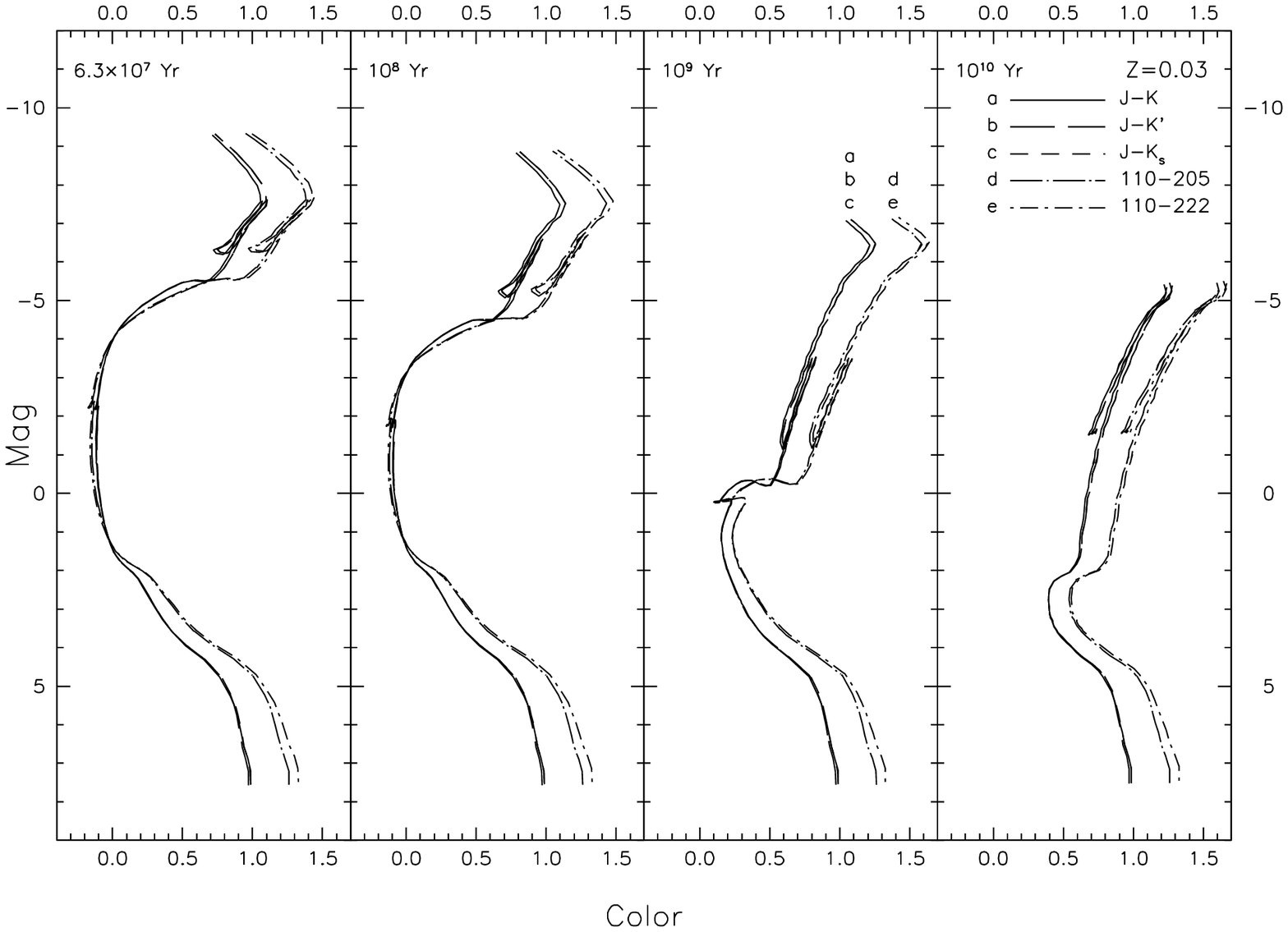}
\caption
{\label{fig:iso4}Same as Figure~\ref{fig:iso1}, but for the $Z=0.03$ model.
Isochrones for $K'$ and $K_s$ are indistinguishable.}
\end{figure}

Isochrones for $A_\lambda = 0$, calculated in this way, are shown in
Figures~\ref{fig:iso1}$-$\ref{fig:iso4} for four different metallicities
and four ages.  The color differences between filters are more prominent
for the highest metallicity isochrones.  In most cases, isochrones for
$K'$ and $K_s$ are nearly indistinguishable, and those for F205W and
F222M are quite close to each other.  In general, for red giants, intrinsic
color differences between the atmospheric and NICMOS filters are 0.2--0.4~mag.

\begin{figure}
\epsscale{1.0}
\plotone{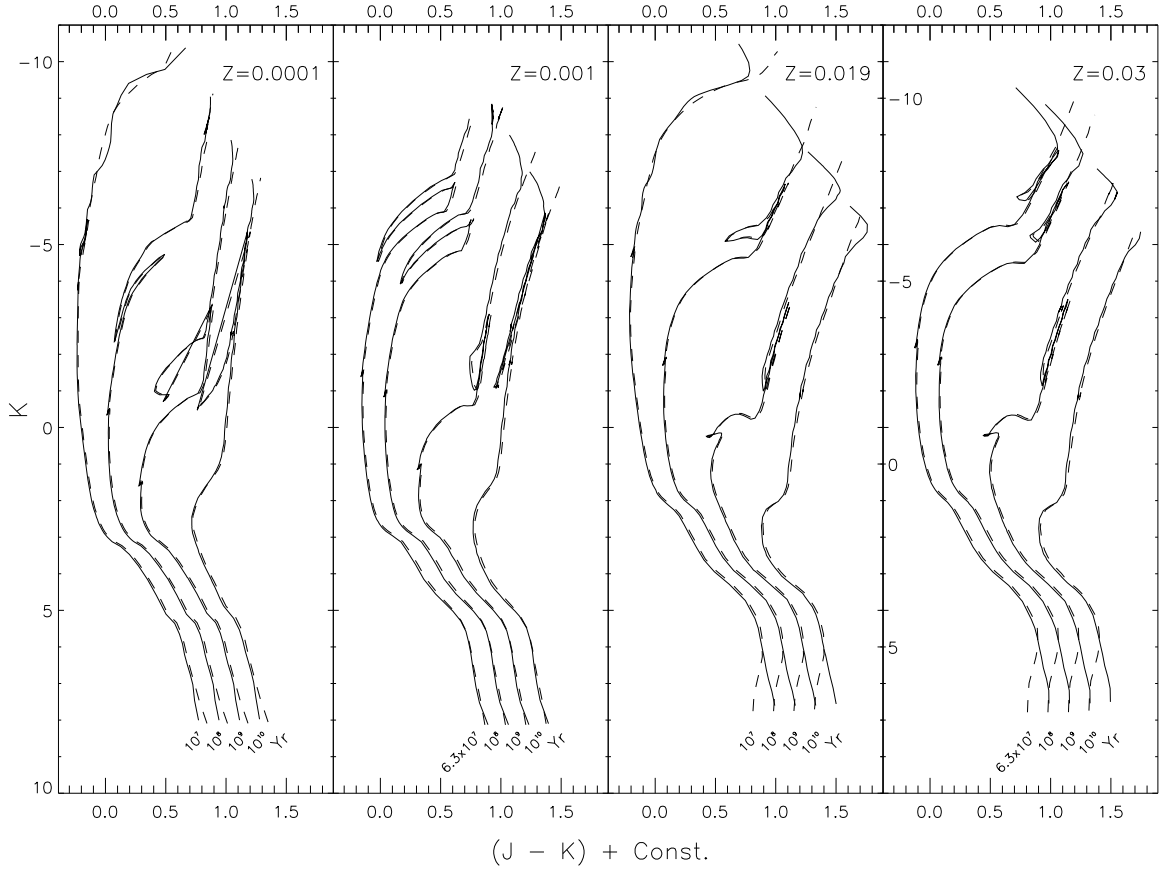}
\caption
{\label{fig:padova}
Plot of $J-K$ vs. $K$ isochrones calculated in the present study ($\it solid
lines$) and those by Girardi et al. (2002; $\it dashed lines$).  Only data
points that have $\log T_{eff} \ge 3500$~K and $\log g \ge 0$ are plotted.
}
\end{figure}

As an independent check of our procedure, in Figure~\ref{fig:padova}
we compare our $J-K$ versus $K$ isochrones to those calculated by
Girardi et al. (2002).  The isochrones match nicely, except
at the extremes.  The discrepancy in the bright end
is caused from the empirical M giant spectra that Girardi et al. (2002)
added to their spectral library, and that in the faint end is by the
addition of late M dwarf spectra.  The discrepancies are considerable only
at the top and bottom $\sim$ 1 mag of the isochrone, where only a
small fraction of giants reside, or else stars are too faint for most
observational situations.

Magnitude transformations between $K$-band filters can be obtained from
our isochrones.  We find that the magnitude difference can be well fitted
by a third-order polynomial for $K < 4$~mag, and by a separate second-order
polynomial for $K > 4$~mag.  The largest residuals from the fit are
0.012~mag for the former and 0.008~mag for the latter.  The coefficients
of the best-fit functions are presented in Tables~\ref{table:trans1} and
\ref{table:trans2}, along with the residuals and fitting ranges.
One useful way of using these tables would be to compare the magnitudes
of helium-burning clump giant stars, which are rather insensitive to
metallicity or age and are often used as distance indicators, observed
with different photometric systems (the clump stars show a small variation
with age, however; see Figer et al. 2004).

\begin{figure}
\epsscale{1.0}
\plotone{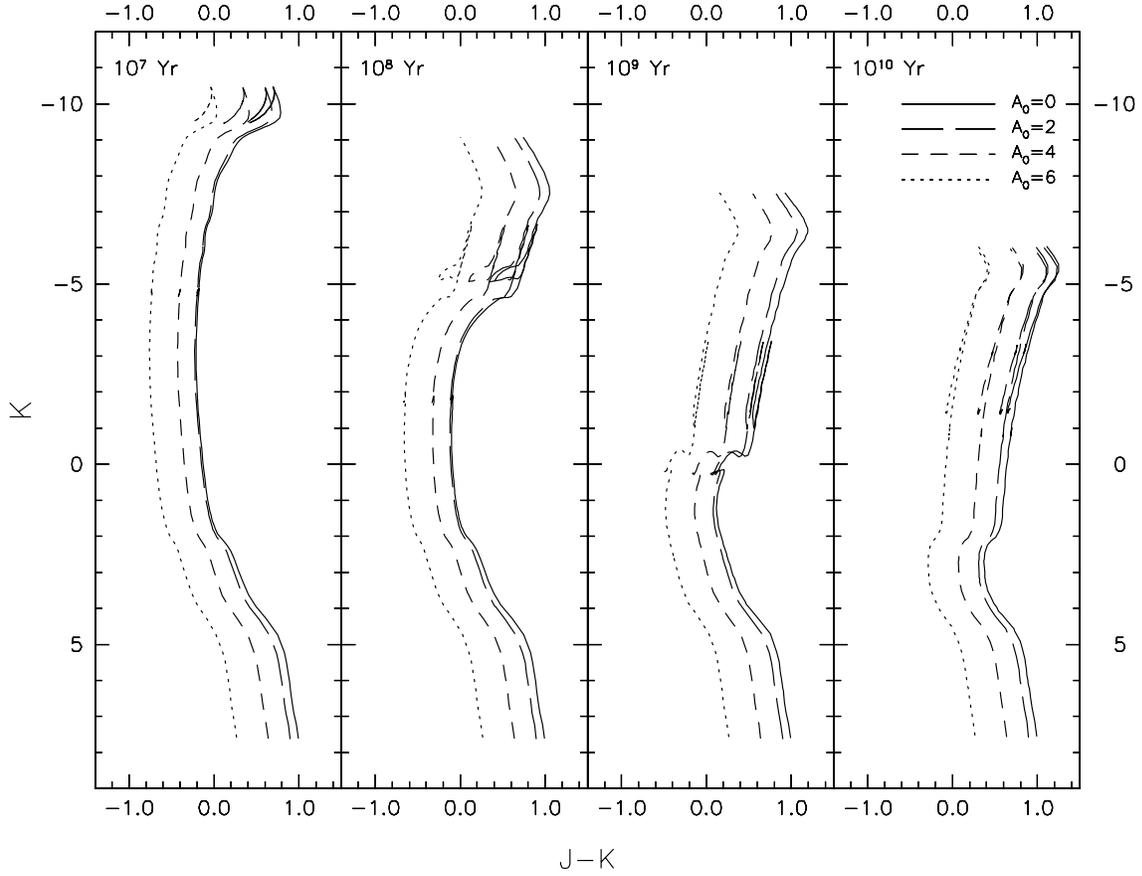}
\caption
{\label{fig:red1}Dereddened $J-K$ vs. $K$ isochrones of the $Z=0.019$ model
with $A_0$=0 ($\it solid line$), $A_0$=2 ($\it long dashed line$), $A_0$=4
($\it short dashed line$), and $A_0$=6 ($\it dotted line$).  The isochrones
are dereddened by an amount $A_0 (\lambda_c /
\lambda_0)^{-1.66}$.  Only data points that have $\log T_{eff} \ge
3500$~K and $\log g \ge 0$ are plotted.}
\end{figure}

\begin{figure}
\epsscale{1.0}
\plotone{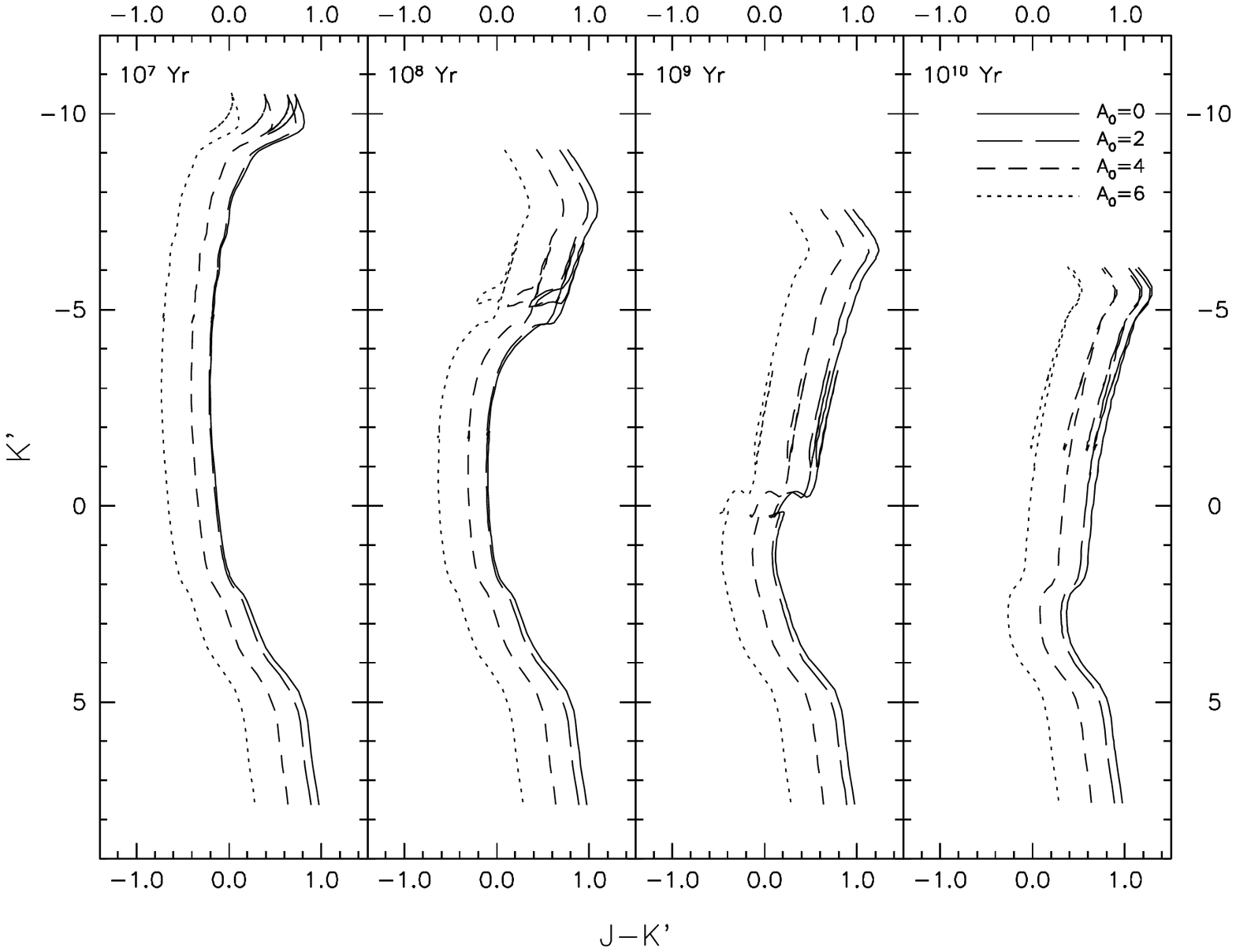}
\caption
{\label{fig:red2}Same as Figure~\ref{fig:red1}, but for $J-K'$ vs. $K'$.}
\end{figure}

\begin{figure}
\epsscale{1.0}
\plotone{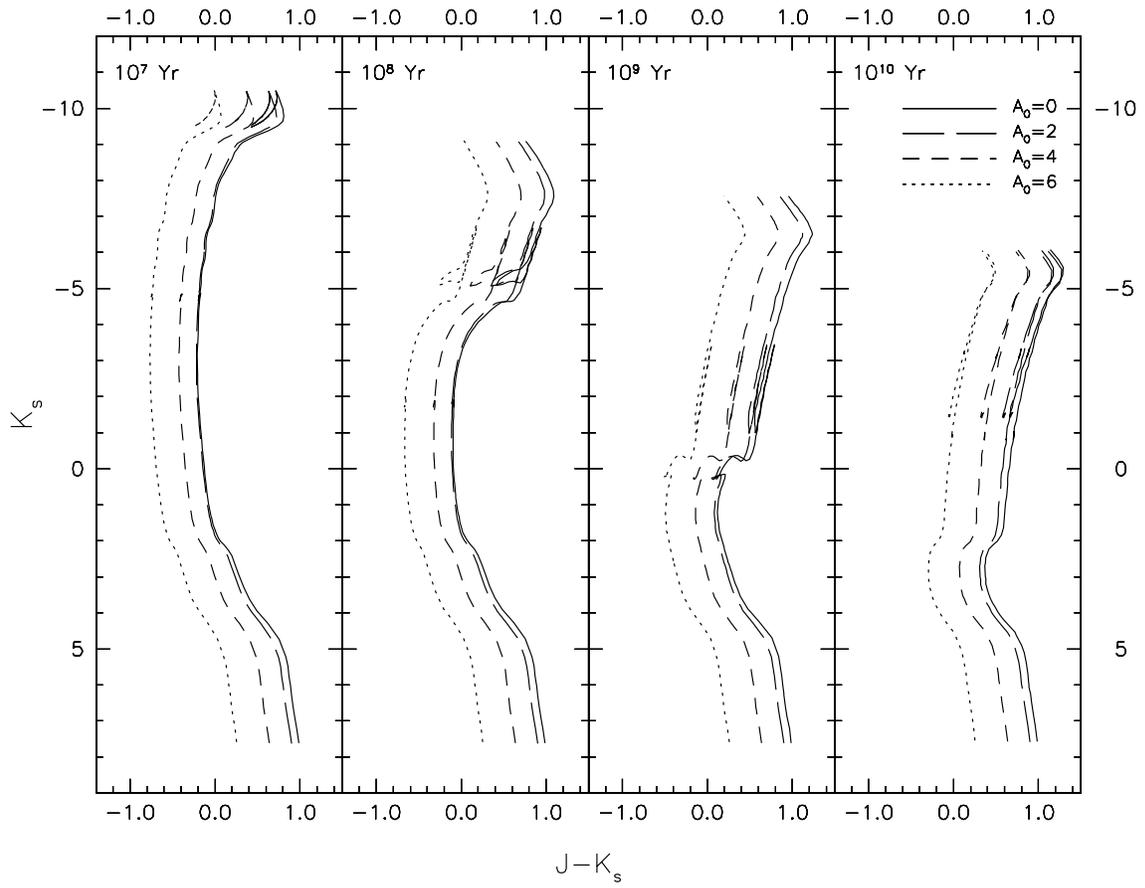}
\caption
{\label{fig:red3}Same as Figure~\ref{fig:red1}, but for $J-K_s$ vs. $K_s$.}
\end{figure}

\begin{figure}
\epsscale{1.0}
\plotone{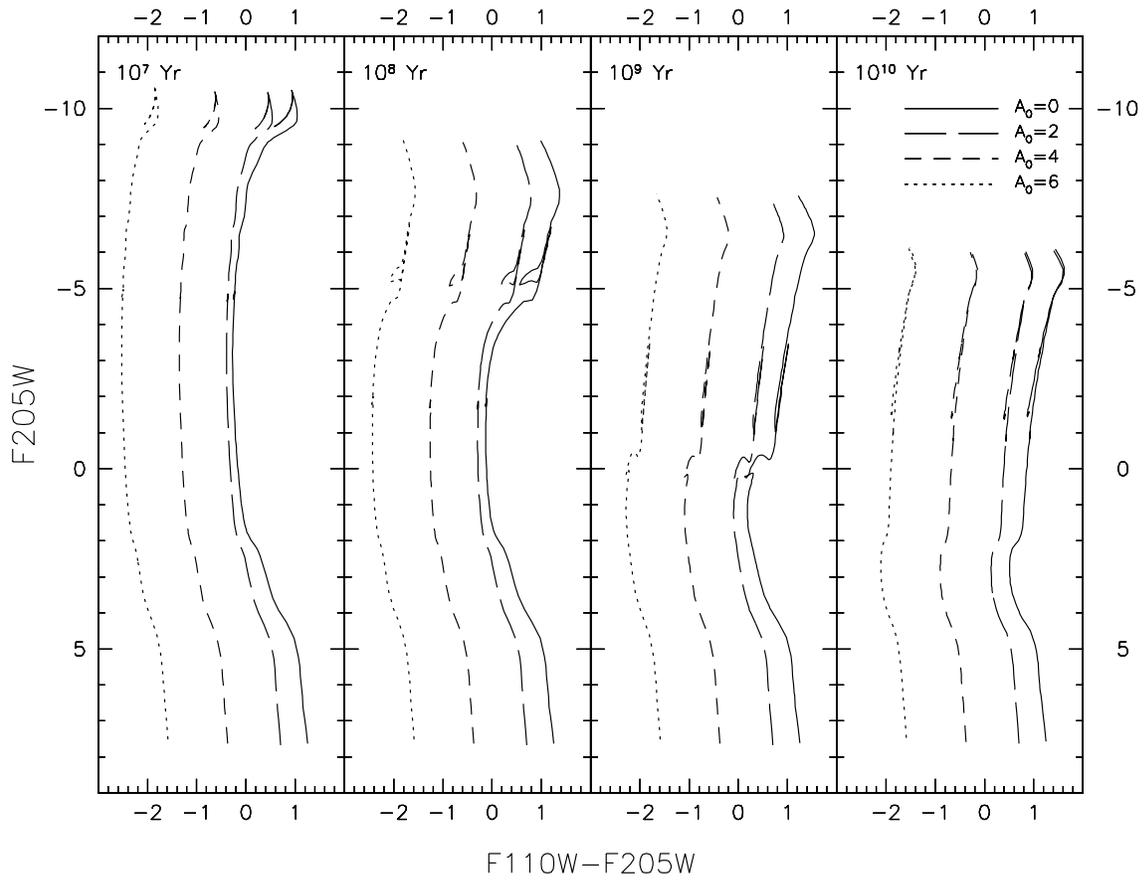}
\caption
{\label{fig:red4}Same as Figure~\ref{fig:red1}, but for F110W$-$F205W vs. F205W.}
\end{figure}

\begin{figure}
\epsscale{1.0}
\plotone{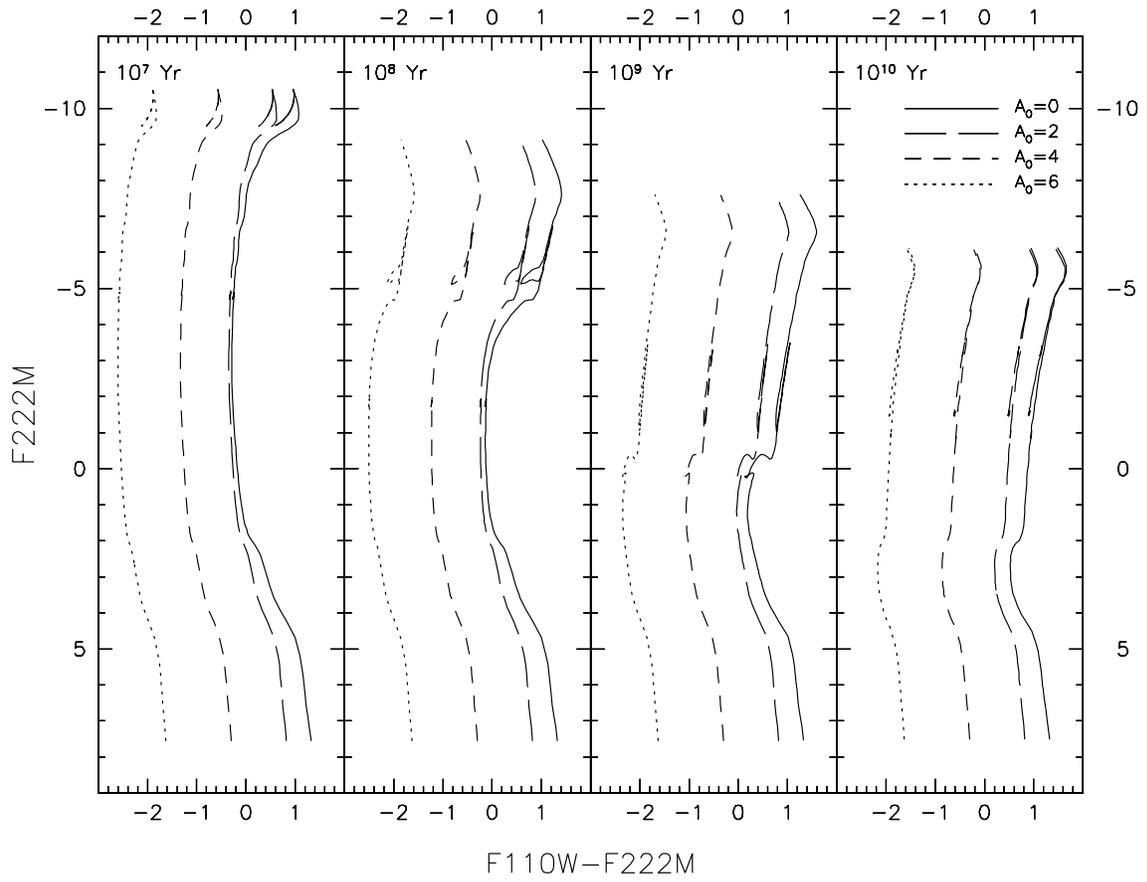}
\caption
{\label{fig:red5}Same as Figure~\ref{fig:red1}, but for F110W$-$F222M vs. F222M.}
\end{figure}

We present here isochrones with $K$-band extinctions of up to 6 mag, some of
which are shown in Figures~\ref{fig:red1}$-$\ref{fig:red5}.  For the
extinction between the $J$ and $K$ bands, we adopt a power law,
\begin{equation}
\label{extinction}
	A_\lambda = A_0 \left ( \frac{\lambda}{\lambda_0}
			\right )^{-\alpha},
\end{equation}
where we choose $\lambda_0 = 2.2 \, \mu$m, and $A_0$ is the extinction at
$\lambda_0$.  When assuming that the transmission functions of the $J$ and $K$
filters are Dirac delta functions centered at 1.24 and 2.21~$\mu$m,
respectively, the extinction law by Rieke et al. (1989) gives
$\alpha=1.59$.
However, as discussed below in this section, the apparent extinction behavior
of isochrones in the CMD can differ from the actual extinction law, as a
result of a nonzero width and asymmetry of the filter transmission functions.
We find that $\alpha=1.66$ makes the isochrone for the $Z = 0.019$,
age = $10^9$~yr model behave in the CMD as if it followed an extinction law
with $\alpha=1.59$.  We choose this particular isochrone for calibrating
the extinction law, with the assumption that the stars used in Rieke et al.
(1989) to derive their extinction law, which are the stars in the central
parsec of our Galaxy, can be represented by the same metallicity and age.
For the sake of comparison, isochrones in
Figures~\ref{fig:red1}$-$\ref{fig:red5} have been dereddened by the amount
$A_0 ( \lambda_c / \lambda_0 )^{-1.66}$, where the central wavelength
of the filter $\lambda_c$ is defined by equation (8) of Paper I, and
given in Table~\ref{table:lambdac}.

Since we have dereddened the isochrones with the known amount of
extinction at $\lambda_c$, all the dereddened isochrones with different
extinction values in Figures~\ref{fig:red1}$-$\ref{fig:red5} should
be coincident if the filter transmission functions were Dirac delta
functions centered at $\lambda_c$.  As in Paper I, the dereddened
isochrones misalign significantly, and this implies that the amount of
extinction inferred from a CMD is sensitively dependent on the shape of
the filter transmission function.

\begin{figure}
\epsscale{1.0}
\plotone{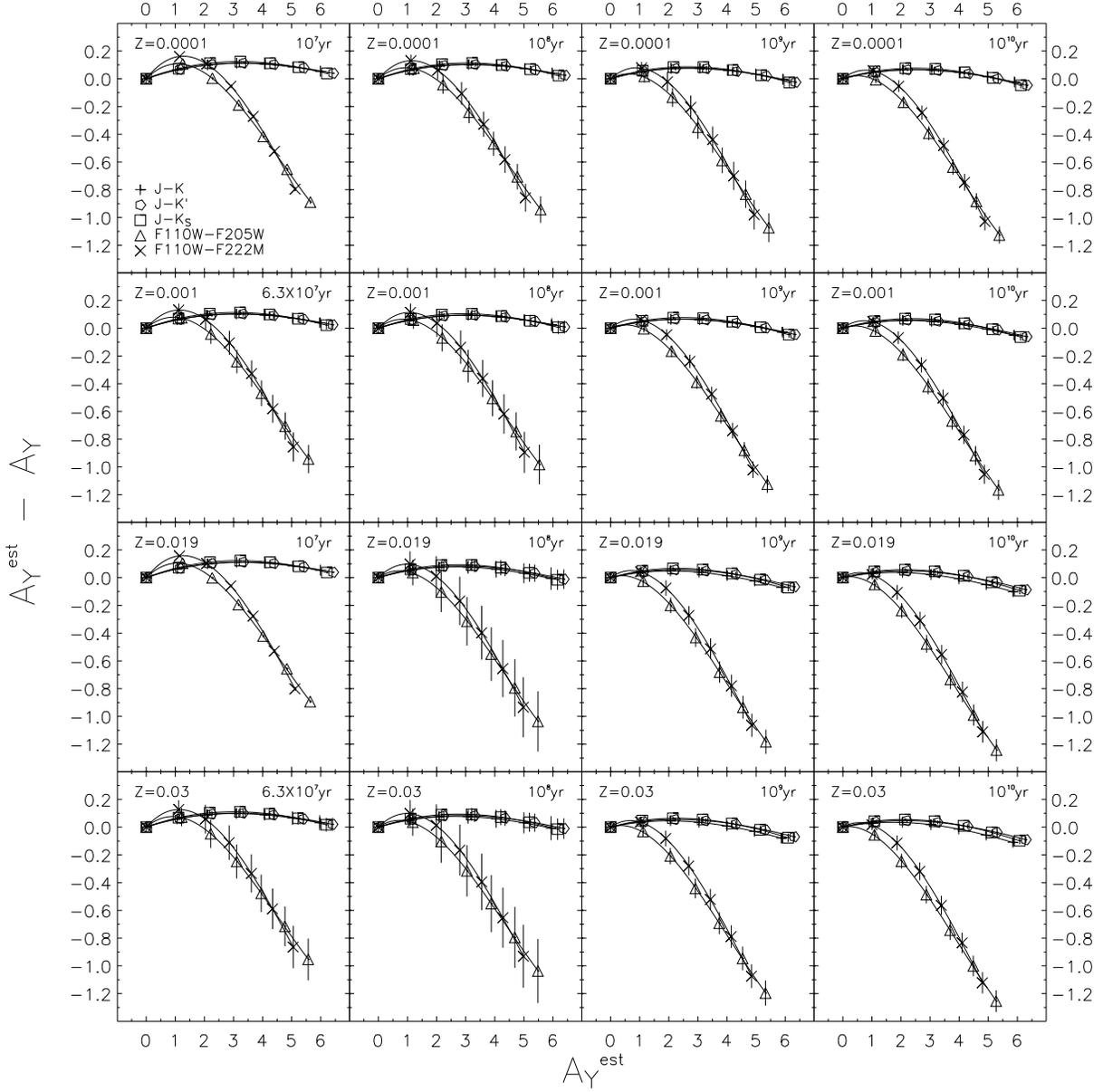}
\caption
{\label{fig:adiff1}Difference between the extinction values
that are estimated by eq.~(\ref{A_est}) using the colors from our
reddened isochrones and the actual extinction values.  A constant
value of 1.59 is used for $\alpha$ in eq.~(\ref{A_est}).
The extinction of each isochrone has been estimated with the mean color
(for $A^{\rm est}_Y$) and the mean magnitude (for $A_Y$) of the reddened
isochrone data points whose intrinsic $K$-band magnitudes are between $-$6
and 0 mag.  The error bar represents the standard deviation of
$A_Y^{\rm est}-A_Y$ values.
}
\end{figure}

When estimating the amount of extinction from an observed CMD, one
converts an observed color excess to an extinction value, following
an assumed extinction law, which usually has the form of a power law.
When one has photometric data from a pair of two filters, $X$ and $Y$, the
amount of extinction can be estimated by
\begin{eqnarray}
\label{A_est}
	A_Y^{est} & = & \frac{(m_X-m_Y)-(m_X-m_Y)_0}{A_X/A_Y-1} \cr
                  & = & \frac{(m_X-m_Y)-(m_X-m_Y)_0}
			     {(\lambda_X/\lambda_Y)^{-\alpha}-1},
\end{eqnarray}
where $m_X$, $m_Y$ and $\lambda_X$, $\lambda_Y$ are the magnitudes and
the central wavelengths of the two filters, respectively, and subscript 0
denotes the intrinsic value.  For estimating extinction from our isochrones,
we first use $\alpha=1.59$.  Figure~\ref{fig:adiff1} shows the difference
between the inferred extinction values, using equation~(\ref{A_est}) and
colors from our reddened isochrones, and the actual extinction values.
Here the extinction of each isochrone has been calculated using the mean
color (for $A^{est}_Y$) and magnitude (for $A_Y$) of the reddened isochrone
data points having intrinsic $K$-band magnitudes between $-$6 and 0 mag.
As the figure shows, the differences between estimated and actual extinction
values are much larger for the NICMOS filter pairs.  The largest
relative difference is $\sim 24$\%, and the largest absolute difference is
1.25 mag.  Note that the extinction estimates for the $Z = 0.019$
and age = $10^9$~yr model inferred from $H$ and $K$ are very close to
the actual extinction values, justifying our choice of $\alpha=1.66$
for equation~(\ref{extinction}).  The error bar in the figure represents
the standard deviation of $A^{est}_Y-A_Y$ values.  Some of the F110W
isochrones show quite large deviations, as pointed out in Appendix A of
Lee et al. (2001).

\begin{figure}
\epsscale{0.6}
\plotone{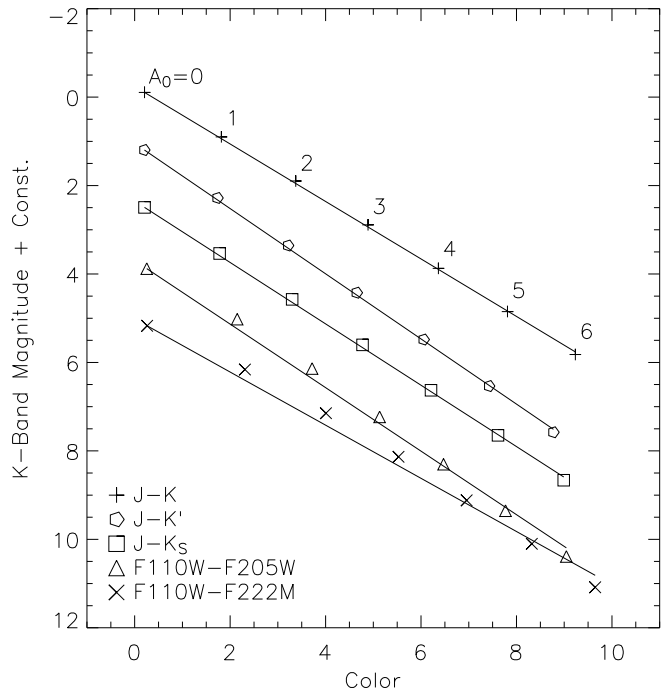}
\caption
{\label{fig:extlaw}Reddened magnitudes of $K$-band filters and reddened
colors for the $Z=0.019$ and age = $10^9$~yr isochrone data point whose
intrinsic $K$ magnitude is 0.  Also shown are the best-fit straight
lines that go through the data point for $A_\lambda = 0$ for each filter pair.
}
\end{figure}

\begin{figure}
\epsscale{1.0}
\plotone{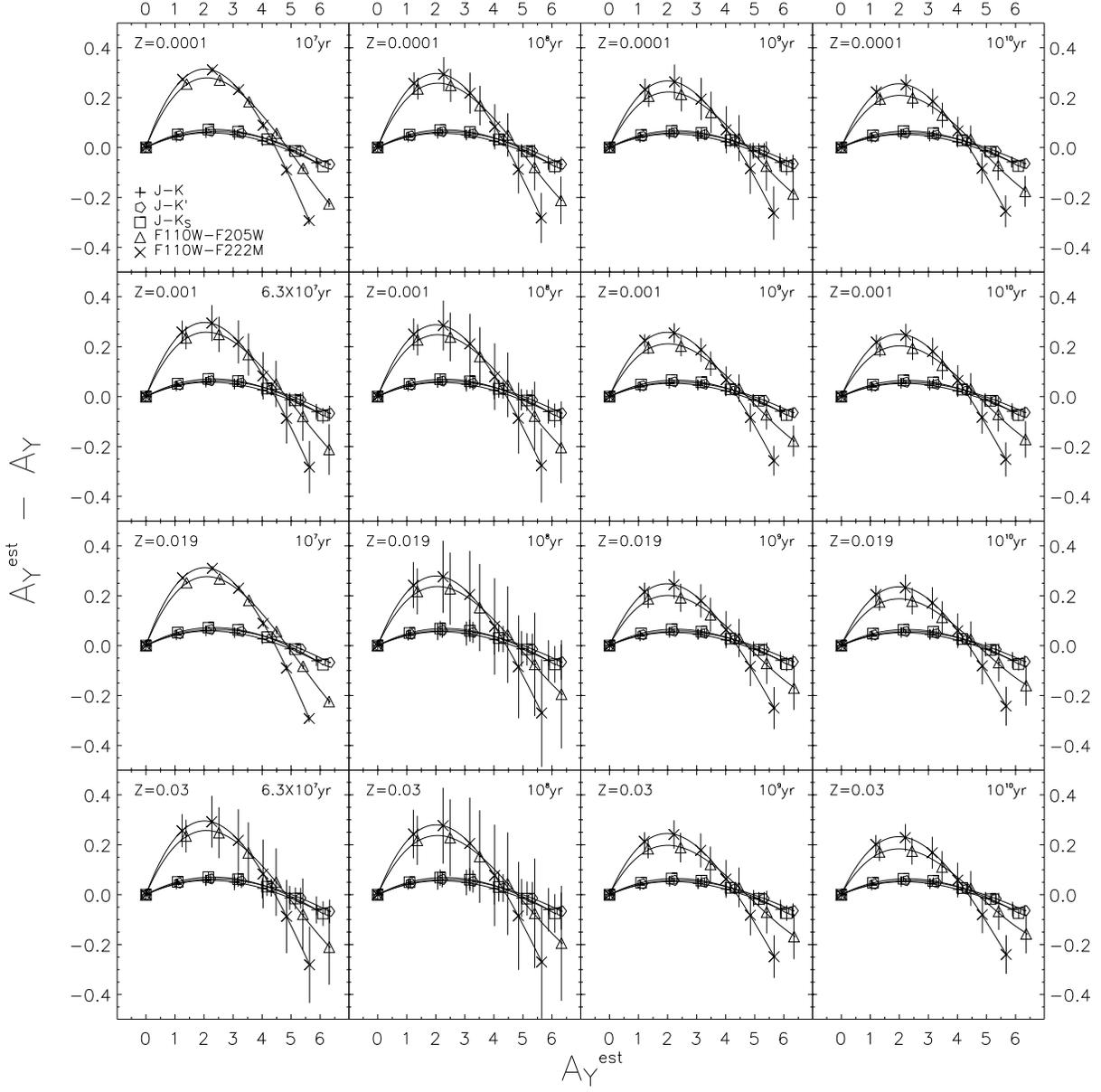}
\caption
{\label{fig:adiff3}Same as Figure~\ref{fig:adiff1}, but using $\alpha_{eff}$
for eq.~(\ref{A_est}).
}
\end{figure}

To reduce the problems seen in Figure~\ref{fig:adiff1}, Paper I introduced
an ``effective extinction slope'' $\alpha_{eff}$ for each filter pair and
isochrone model, which is defined such that it better describes the
extinction behavior in the CMD:
\begin{equation}
\label{alpha_eff}
	\alpha_{eff} = - \frac{ \log (1+1/b) }{ \log (\lambda_X/\lambda_Y) },
\end{equation}
where $b$ is the slope of the straight line that fits the distribution of
reddened magnitudes versus reddened colors, as in Figure~\ref{fig:extlaw}.
This figure shows reddened $K$-band magnitudes and colors for the $Z = 0.019$
and age = $10^9$~yr isochrone (the figure only shows an isochrone data
point whose intrinsic $K$ magnitude is 0, as an example).
We calculate $b$ for data points of each isochrone whose intrinsic
$K$ magnitudes are between $-6$ and 0~mag, and take an average for
each isochrone model.  Table~\ref{table:alpha_eff} shows the averages
and standard deviations of $\alpha_{eff}$ values for each isochrone model.
For atmospheric filters, the standard deviations of $\alpha_{eff}$ in an
isochrone is generally much smaller than the differences of average
$\alpha_{eff}$ values between different isochrones, while those for
NICMOS filters are relatively larger.
The average $\alpha_{eff}$ values range from 1.403 to 1.610,
which are 15\% to 0.02\% smaller than the original $\alpha$ value we
adopted for extinction, 1.66.
As seen in Figure~\ref{fig:adiff3}, extinction values estimated by
equation~(\ref{A_est}) with $\alpha_{eff}$ are closer to
the actual values for atmospheric filters, but still deviate significantly
from the actual values for NICMOS filters, because of the nonlinear
extinction seen in Figure~\ref{fig:extlaw}.

\begin{figure}
\epsscale{1.0}
\plotone{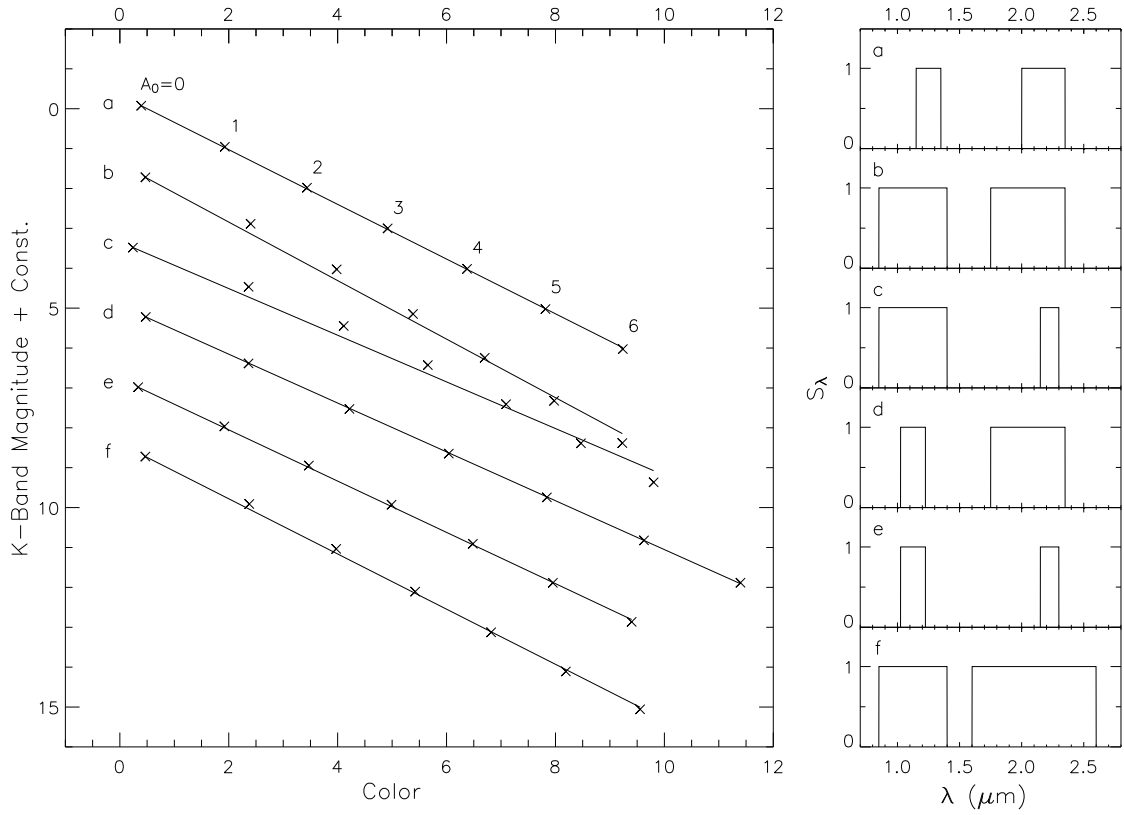}
\caption
{\label{fig:nonlin}Reddened magnitudes and colors ($\it crosses$) for the
$Z=0.019$ and age = $10^9$~yr isochrone data point whose intrinsic $K$
magnitude is 0, for six imaginary filter pairs whose transmission functions
are shown in the right panel.  Also shown are the best-fit straight lines
that go through the data point for $A_\lambda = 0$ for each filter pair.
}
\end{figure}

As pointed out in Paper I, the nonlinear extinction behavior of NICMOS
filters is due to a significant difference in relative widths of the two
filters:  the width to central wavelength ratio $\Delta \lambda / \lambda_c$
is $\sim 0.5$ for F110W, while those for F205W and F222M are $\sim 0.3$
and $\sim 0.07$, respectively.
Figure~\ref{fig:nonlin} shows the effect of the filter width by comparing
the extinction behavior of six imaginary filter pairs.  Filter pair $a$
represents $J$ and $K$, whose $\Delta \lambda / \lambda_c$ values are
both $\sim 0.16$, and its extinction behavior in the CMD is nearly linear.
On the other hand, filter pairs $b$ and $c$, which represent filter pairs
F110W--F205W and F110W--F222M, show considerable nonlinearity.  When
the $\Delta \lambda / \lambda_c$ of the short-wavelength filter is reduced
by $\sim 60$\%, however, the extinction behaves much more linearly
($d$ and $e$).  This shows that the nonlinear extinction in
filter pairs F110W--F205W and F110W--F222M is due to a relatively larger
$\Delta \lambda / \lambda_c$ value of the F110W filter.  When both
filters have the same large $\Delta \lambda / \lambda_c$ values
($\sim 0.5$), the extinction becomes almost linear again ($f$).

The introduction of effective extinction slopes does not alleviate the
nonlinear extinction problem of NICMOS filter pairs.
So in Table~\ref{table:nonlin} we provide the coefficients of best-fit
third-order polynomials of the extinction curves for NICMOS
filter pairs shown in Figure~\ref{fig:adiff1} so that one can accurately
estimate the extinction value for NICMOS filter
pairs as well.  Note that we assumed $\alpha = 1.59$ for all isochrone models
when estimating $A_Y^{est}$ in Figure~\ref{fig:adiff1}.

As in paper I, we find that for the filters whose extinction behavior
is relatively linear, the transformation of extinction values from filter
$Y$ to filter $Y'$ can be obtained by
\begin{equation}
\label{A_trans}
	A_{Y'} = A_Y \left ( \frac{\lambda_{Y'}}{\lambda_Y} \right )^{-\alpha},
\end{equation}
if $A_Y$ is estimated with $\alpha_{eff}$, and the original $\alpha$ value
of 1.66 is used in the above equation.

\section{SUMMARY}
\label{sec:summary}

We have calculated in a consistent way five near-infrared theoretical
isochrones for filter pairs composed of $J$ and $K$ filters: $J$--$K$,
$J$--$K'$, $J$--$K_s$, F110W--F205W, and F110W--F222M.  We presented
isochrones for a $Z$ of 0.0001--0.03 and an age of $10^7$--$10^{10}$~yr.
Even in the same Vega magnitude system, near-infrared colors of the same
isochrone can be different by up to $\sim 0.4$ mag at the bright end of
the isochrone for different filter pairs.
The difference in intrinsic colors for a red giant for atmospheric
filters and the {\it HST} NICMOS filters is generally 0.2--0.4 mag.
We have provided magnitude transformations between $K$-band filters
as a function of color from $J$ and $K$ band filters.
We also presented isochrones with $A_K$ of up to 6 mag.
We found that care is needed when comparing extinction
values that are estimated using different filter pairs, in particular
when comparing those of atmospheric and NICMOS filter pairs:
extinction values inferred using NICMOS filters can be in error 
by up to 1.3 mag.  To alleviate this problem, we introduced an
``effective extinction slope'' for each filter pair and isochrone model,
which describes the extinction-dependent behavior of isochrones in the
observed CMD.  We also provided a procedure to accurately estimate
the extinction value for NICMOS filter pairs, whose extinction curves
in the CMD are highly nonlinear.

\acknowledgements
We thank Jae-Woo Lee for a helpful discussion.
S. S. K. was supported by the Astrophysical Research Center for the
Structure and Evolution of the Cosmos (ARCSEC) of the Korea Science and
Engineering Foundation through the Science Research Center (SRC) program.
M. G. L. was in part supported by the ABRL (R14-2002-058-01000-0) and the BK21
program.


\clearpage
\begin{deluxetable}{cclrrrrcc}
\tabletypesize{\scriptsize}
\tablecolumns{9}
\tablewidth{0pt}
\tablecaption{
\label{table:trans1}Best-Fit Coefficients for Magnitude
Differences ($K < 4$~mag)}
\tablehead{
\colhead{} &
\colhead{Magnitude} &
\colhead{} &
\colhead{} &
\colhead{} &
\colhead{} &
\colhead{} &
\colhead{Residual\tablenotemark{a}} &
\colhead{Fitting Range\tablenotemark{b}} \\
\colhead{Color} &
\colhead{Difference} &
\colhead{$Z$} &
\colhead{$c_0$} &
\colhead{$c_1$} &
\colhead{$c_2$} &
\colhead{$c_3$} &
\colhead{(mag)} &
\colhead{(mag $\sim$ mag)}
}
\startdata
$J - K$  & $K'$ $-$$K$  & 0.0001 & $-0.001$ & $ 0.028$ & $-0.064$ & $ 0.057$ & $0.004$ & $-0.236 \sim 0.720$ \\
$J - K$  & $K'$ $-$$K$  & 0.001  & $-0.001$ & $ 0.032$ & $-0.039$ & $-0.036$ & $0.007$ & $-0.162 \sim 0.859$ \\
$J - K$  & $K'$ $-$$K$  & 0.019  & $ 0.002$ & $ 0.036$ & $-0.147$ & $ 0.074$ & $0.011$ & $-0.213 \sim 1.250$ \\
$J - K$  & $K'$ $-$$K$  & 0.03   & $ 0.001$ & $ 0.042$ & $-0.163$ & $ 0.083$ & $0.007$ & $-0.129 \sim 1.237$ \\
$J - K$  & $K_s$$-$$K$  & 0.0001 & $-0.000$ & $ 0.012$ & $-0.013$ & $ 0.006$ & $0.002$ & $-0.236 \sim 0.720$ \\
$J - K$  & $K_s$$-$$K$  & 0.001  & $-0.000$ & $ 0.015$ & $-0.003$ & $-0.052$ & $0.006$ & $-0.162 \sim 0.859$ \\
$J - K$  & $K_s$$-$$K$  & 0.019  & $ 0.002$ & $ 0.017$ & $-0.103$ & $ 0.051$ & $0.009$ & $-0.213 \sim 1.250$ \\
$J - K$  & $K_s$$-$$K$  & 0.03   & $ 0.001$ & $ 0.024$ & $-0.121$ & $ 0.061$ & $0.006$ & $-0.129 \sim 1.237$ \\
$J - K$  & F205W$-$$K$  & 0.0001 & $-0.030$ & $ 0.052$ & $-0.143$ & $ 0.150$ & $0.006$ & $-0.236 \sim 0.720$ \\
$J - K$  & F205W$-$$K$  & 0.001  & $-0.030$ & $ 0.056$ & $-0.095$ & $ 0.028$ & $0.005$ & $-0.162 \sim 0.859$ \\
$J - K$  & F205W$-$$K$  & 0.019  & $-0.029$ & $ 0.060$ & $-0.147$ & $ 0.074$ & $0.007$ & $-0.213 \sim 1.250$ \\
$J - K$  & F205W$-$$K$  & 0.03   & $-0.029$ & $ 0.061$ & $-0.147$ & $ 0.077$ & $0.004$ & $-0.129 \sim 1.237$ \\
$J - K$  & F222M$-$$K$  & 0.0001 & $-0.031$ & $-0.002$ & $-0.001$ & $-0.019$ & $0.005$ & $-0.236 \sim 0.720$ \\
$J - K$  & F222M$-$$K$  & 0.001  & $-0.030$ & $ 0.001$ & $-0.010$ & $-0.046$ & $0.006$ & $-0.162 \sim 0.859$ \\
$J - K$  & F222M$-$$K$  & 0.019  & $-0.028$ & $ 0.002$ & $-0.113$ & $ 0.060$ & $0.012$ & $-0.213 \sim 1.250$ \\
$J - K$  & F222M$-$$K$  & 0.03   & $-0.028$ & $ 0.009$ & $-0.134$ & $ 0.071$ & $0.007$ & $-0.129 \sim 1.237$ \\
$J - K'$  & $K$ $-$$K'$  & 0.0001 & $ 0.001$ & $-0.029$ & $ 0.068$ & $-0.061$ & $0.004$ & $-0.224 \sim 0.715$ \\
$J - K'$  & $K$ $-$$K'$  & 0.001  & $ 0.001$ & $-0.033$ & $ 0.047$ & $ 0.024$ & $0.006$ & $-0.155 \sim 0.881$ \\
$J - K'$  & $K$ $-$$K'$  & 0.019  & $-0.001$ & $-0.037$ & $ 0.142$ & $-0.070$ & $0.011$ & $-0.203 \sim 1.290$ \\
$J - K'$  & $K$ $-$$K'$  & 0.03   & $-0.001$ & $-0.042$ & $ 0.154$ & $-0.076$ & $0.006$ & $-0.123 \sim 1.278$ \\
$J - K'$  & $K_s$$-$$K'$  & 0.0001 & $ 0.001$ & $-0.017$ & $ 0.054$ & $-0.055$ & $0.003$ & $-0.224 \sim 0.715$ \\
$J - K'$  & $K_s$$-$$K'$  & 0.001  & $ 0.001$ & $-0.018$ & $ 0.038$ & $-0.017$ & $0.002$ & $-0.155 \sim 0.881$ \\
$J - K'$  & $K_s$$-$$K'$  & 0.019  & $ 0.000$ & $-0.019$ & $ 0.043$ & $-0.022$ & $0.003$ & $-0.203 \sim 1.290$ \\
$J - K'$  & $K_s$$-$$K'$  & 0.03   & $ 0.001$ & $-0.019$ & $ 0.040$ & $-0.021$ & $0.002$ & $-0.123 \sim 1.278$ \\
$J - K'$  & F205W$-$$K'$  & 0.0001 & $-0.029$ & $ 0.025$ & $-0.085$ & $ 0.099$ & $0.003$ & $-0.224 \sim 0.715$ \\
$J - K'$  & F205W$-$$K'$  & 0.001  & $-0.029$ & $ 0.024$ & $-0.054$ & $ 0.058$ & $0.004$ & $-0.155 \sim 0.881$ \\
$J - K'$  & F205W$-$$K'$  & 0.019  & $-0.031$ & $ 0.024$ & $-0.002$ & $ 0.001$ & $0.005$ & $-0.203 \sim 1.290$ \\
$J - K'$  & F205W$-$$K'$  & 0.03   & $-0.030$ & $ 0.019$ & $ 0.012$ & $-0.004$ & $0.005$ & $-0.123 \sim 1.278$ \\
$J - K'$  & F222M$-$$K'$  & 0.0001 & $-0.030$ & $-0.032$ & $ 0.067$ & $-0.080$ & $0.005$ & $-0.224 \sim 0.715$ \\
$J - K'$  & F222M$-$$K'$  & 0.001  & $-0.029$ & $-0.032$ & $ 0.032$ & $-0.012$ & $0.003$ & $-0.155 \sim 0.881$ \\
$J - K'$  & F222M$-$$K'$  & 0.019  & $-0.030$ & $-0.035$ & $ 0.036$ & $-0.015$ & $0.004$ & $-0.203 \sim 1.290$ \\
$J - K'$  & F222M$-$$K'$  & 0.03   & $-0.029$ & $-0.035$ & $ 0.031$ & $-0.013$ & $0.003$ & $-0.123 \sim 1.278$ \\
$J - K_s$ & $K$ $-$$K_s$ & 0.0001 & $ 0.000$ & $-0.012$ & $ 0.013$ & $-0.005$ & $0.002$ & $-0.232 \sim 0.718$ \\
$J - K_s$ & $K$ $-$$K_s$ & 0.001  & $ 0.000$ & $-0.015$ & $ 0.008$ & $ 0.043$ & $0.006$ & $-0.160 \sim 0.879$ \\
$J - K_s$ & $K$ $-$$K_s$ & 0.019  & $-0.002$ & $-0.017$ & $ 0.098$ & $-0.047$ & $0.009$ & $-0.209 \sim 1.289$ \\
$J - K_s$ & $K$ $-$$K_s$ & 0.03   & $-0.001$ & $-0.023$ & $ 0.113$ & $-0.055$ & $0.006$ & $-0.127 \sim 1.278$ \\
$J - K_s$ & $K'$ $-$$K_s$ & 0.0001 & $-0.001$ & $ 0.017$ & $-0.052$ & $ 0.053$ & $0.003$ & $-0.232 \sim 0.718$ \\
$J - K_s$ & $K'$ $-$$K_s$ & 0.001  & $-0.001$ & $ 0.018$ & $-0.037$ & $ 0.017$ & $0.002$ & $-0.160 \sim 0.879$ \\
$J - K_s$ & $K'$ $-$$K_s$ & 0.019  & $-0.000$ & $ 0.019$ & $-0.042$ & $ 0.022$ & $0.003$ & $-0.209 \sim 1.289$ \\
$J - K_s$ & $K'$ $-$$K_s$ & 0.03   & $-0.001$ & $ 0.018$ & $-0.039$ & $ 0.021$ & $0.002$ & $-0.127 \sim 1.278$ \\
$J - K_s$ & F205W$-$$K_s$ & 0.0001 & $-0.029$ & $ 0.041$ & $-0.133$ & $ 0.147$ & $0.005$ & $-0.232 \sim 0.718$ \\
$J - K_s$ & F205W$-$$K_s$ & 0.001  & $-0.030$ & $ 0.042$ & $-0.090$ & $ 0.075$ & $0.004$ & $-0.160 \sim 0.879$ \\
$J - K_s$ & F205W$-$$K_s$ & 0.019  & $-0.031$ & $ 0.042$ & $-0.044$ & $ 0.022$ & $0.006$ & $-0.209 \sim 1.289$ \\
$J - K_s$ & F205W$-$$K_s$ & 0.03   & $-0.031$ & $ 0.037$ & $-0.027$ & $ 0.016$ & $0.005$ & $-0.127 \sim 1.278$ \\
$J - K_s$ & F222M$-$$K_s$ & 0.0001 & $-0.030$ & $-0.014$ & $ 0.012$ & $-0.024$ & $0.005$ & $-0.232 \sim 0.718$ \\
$J - K_s$ & F222M$-$$K_s$ & 0.001  & $-0.030$ & $-0.014$ & $-0.007$ & $ 0.006$ & $0.004$ & $-0.160 \sim 0.879$ \\
$J - K_s$ & F222M$-$$K_s$ & 0.019  & $-0.030$ & $-0.016$ & $-0.008$ & $ 0.008$ & $0.005$ & $-0.209 \sim 1.289$ \\
$J - K_s$ & F222M$-$$K_s$ & 0.03   & $-0.030$ & $-0.016$ & $-0.010$ & $ 0.008$ & $0.004$ & $-0.127 \sim 1.278$ \\
F110W$-$F205W & $K$ $-$F205W & 0.0001 & $ 0.030$ & $-0.040$ & $ 0.085$ & $-0.070$ & $0.007$ & $-0.304 \sim 0.928$ \\
F110W$-$F205W & $K$ $-$F205W & 0.001  & $ 0.030$ & $-0.043$ & $ 0.053$ & $-0.010$ & $0.005$ & $-0.210 \sim 1.122$ \\
F110W$-$F205W & $K$ $-$F205W & 0.019  & $ 0.029$ & $-0.047$ & $ 0.086$ & $-0.033$ & $0.006$ & $-0.272 \sim 1.621$ \\
F110W$-$F205W & $K$ $-$F205W & 0.03   & $ 0.029$ & $-0.047$ & $ 0.086$ & $-0.034$ & $0.004$ & $-0.166 \sim 1.615$ \\
F110W$-$F205W & $K'$ $-$F205W & 0.0001 & $ 0.029$ & $-0.018$ & $ 0.048$ & $-0.043$ & $0.003$ & $-0.304 \sim 0.928$ \\
F110W$-$F205W & $K'$ $-$F205W & 0.001  & $ 0.029$ & $-0.018$ & $ 0.035$ & $-0.030$ & $0.003$ & $-0.210 \sim 1.122$ \\
F110W$-$F205W & $K'$ $-$F205W & 0.019  & $ 0.031$ & $-0.017$ & $-0.000$ & $-0.000$ & $0.005$ & $-0.272 \sim 1.621$ \\
F110W$-$F205W & $K'$ $-$F205W & 0.03   & $ 0.030$ & $-0.013$ & $-0.010$ & $ 0.003$ & $0.005$ & $-0.166 \sim 1.615$ \\
F110W$-$F205W & $K_s$$-$F205W & 0.0001 & $ 0.029$ & $-0.031$ & $ 0.078$ & $-0.067$ & $0.005$ & $-0.304 \sim 0.928$ \\
F110W$-$F205W & $K_s$$-$F205W & 0.001  & $ 0.030$ & $-0.032$ & $ 0.055$ & $-0.036$ & $0.004$ & $-0.210 \sim 1.122$ \\
F110W$-$F205W & $K_s$$-$F205W & 0.019  & $ 0.031$ & $-0.032$ & $ 0.026$ & $-0.011$ & $0.006$ & $-0.272 \sim 1.621$ \\
F110W$-$F205W & $K_s$$-$F205W & 0.03   & $ 0.031$ & $-0.027$ & $ 0.014$ & $-0.007$ & $0.005$ & $-0.166 \sim 1.615$ \\
F110W$-$F205W & F222M$-$F205W & 0.0001 & $-0.001$ & $-0.041$ & $ 0.085$ & $-0.079$ & $0.007$ & $-0.304 \sim 0.928$ \\
F110W$-$F205W & F222M$-$F205W & 0.001  & $-0.000$ & $-0.042$ & $ 0.050$ & $-0.033$ & $0.004$ & $-0.210 \sim 1.122$ \\
F110W$-$F205W & F222M$-$F205W & 0.019  & $ 0.001$ & $-0.043$ & $ 0.018$ & $-0.006$ & $0.006$ & $-0.272 \sim 1.621$ \\
F110W$-$F205W & F222M$-$F205W & 0.03   & $ 0.001$ & $-0.037$ & $ 0.005$ & $-0.002$ & $0.004$ & $-0.166 \sim 1.615$ \\
F110W$-$F222M & $K$ $-$F222M & 0.0001 & $ 0.031$ & $ 0.001$ & $ 0.001$ & $ 0.008$ & $0.005$ & $-0.324 \sim 0.955$ \\
F110W$-$F222M & $K$ $-$F222M & 0.001  & $ 0.030$ & $-0.001$ & $ 0.004$ & $ 0.021$ & $0.005$ & $-0.222 \sim 1.152$ \\
F110W$-$F222M & $K$ $-$F222M & 0.019  & $ 0.028$ & $-0.003$ & $ 0.063$ & $-0.024$ & $0.011$ & $-0.291 \sim 1.665$ \\
F110W$-$F222M & $K$ $-$F222M & 0.03   & $ 0.028$ & $-0.009$ & $ 0.076$ & $-0.029$ & $0.006$ & $-0.176 \sim 1.669$ \\
F110W$-$F222M & $K'$ $-$F222M & 0.0001 & $ 0.030$ & $ 0.021$ & $-0.033$ & $ 0.032$ & $0.005$ & $-0.324 \sim 0.955$ \\
F110W$-$F222M & $K'$ $-$F222M & 0.001  & $ 0.029$ & $ 0.022$ & $-0.013$ & $ 0.003$ & $0.003$ & $-0.222 \sim 1.152$ \\
F110W$-$F222M & $K'$ $-$F222M & 0.019  & $ 0.030$ & $ 0.024$ & $-0.017$ & $ 0.005$ & $0.004$ & $-0.291 \sim 1.665$ \\
F110W$-$F222M & $K'$ $-$F222M & 0.03   & $ 0.029$ & $ 0.024$ & $-0.014$ & $ 0.004$ & $0.003$ & $-0.176 \sim 1.669$ \\
F110W$-$F222M & $K_s$$-$F222M & 0.0001 & $ 0.030$ & $ 0.010$ & $-0.006$ & $ 0.011$ & $0.005$ & $-0.324 \sim 0.955$ \\
F110W$-$F222M & $K_s$$-$F222M & 0.001  & $ 0.030$ & $ 0.009$ & $ 0.005$ & $-0.003$ & $0.004$ & $-0.222 \sim 1.152$ \\
F110W$-$F222M & $K_s$$-$F222M & 0.019  & $ 0.030$ & $ 0.011$ & $ 0.007$ & $-0.004$ & $0.005$ & $-0.291 \sim 1.665$ \\
F110W$-$F222M & $K_s$$-$F222M & 0.03   & $ 0.030$ & $ 0.010$ & $ 0.009$ & $-0.005$ & $0.004$ & $-0.176 \sim 1.669$ \\
F110W$-$F222M & F205W$-$F222M & 0.0001 & $ 0.001$ & $ 0.038$ & $-0.076$ & $ 0.070$ & $0.007$ & $-0.324 \sim 0.955$ \\
F110W$-$F222M & F205W$-$F222M & 0.001  & $ 0.000$ & $ 0.040$ & $-0.045$ & $ 0.030$ & $0.004$ & $-0.222 \sim 1.152$ \\
F110W$-$F222M & F205W$-$F222M & 0.019  & $-0.001$ & $ 0.041$ & $-0.017$ & $ 0.005$ & $0.006$ & $-0.291 \sim 1.665$ \\
F110W$-$F222M & F205W$-$F222M & 0.03   & $-0.001$ & $ 0.036$ & $-0.004$ & $ 0.002$ & $0.004$ & $-0.176 \sim 1.669$ \\
\enddata
\tablecomments{Magnitude differences are fitted to a function
$[{\rm Mag\, Diff}] = c_0 + c_1[{\rm Color}] + c_2[{\rm Color}]^2 + 
c_3[{\rm Color}]^3$.  Only the data points that have $\log T_{eff} \ge
3500$~K and $\log g \ge 0$ were considered for the fitting.}
\tablenotetext{a}{The largest absolute residual.}
\tablenotetext{b}{Color range where the fit is valid.}
\end{deluxetable}

\clearpage
\begin{deluxetable}{cclrrrcc}
\tabletypesize{\scriptsize}
\tablecolumns{8}
\tablewidth{0pt}
\tablecaption{
\label{table:trans2}Best-Fit Coefficients for Magnitude
Differences ($K > 4$~mag)}
\tablehead{
\colhead{} &
\colhead{Magnitude} &
\colhead{} &
\colhead{} &
\colhead{} &
\colhead{} &
\colhead{Residual\tablenotemark{a}} &
\colhead{Fitting Range\tablenotemark{b}} \\
\colhead{Color} &
\colhead{Difference} &
\colhead{$Z$} &
\colhead{$c_0$} &
\colhead{$c_1$} &
\colhead{$c_2$} &
\colhead{(mag)} &
\colhead{(mag $\sim$ mag)}
}
\startdata
$J - K$  & $K'$ $-$$K$  & 0.0001 & $-0.012$ & $ 0.047$ & $-0.003$ & $0.001$ & $ 0.332 \sim 0.772$ \\
$J - K$  & $K'$ $-$$K$  & 0.001  & $ 0.034$ & $-0.126$ & $ 0.127$ & $0.003$ & $ 0.338 \sim 0.893$ \\
$J - K$  & $K'$ $-$$K$  & 0.019  & $ 0.102$ & $-0.331$ & $ 0.252$ & $0.004$ & $ 0.539 \sim 0.992$ \\
$J - K$  & $K'$ $-$$K$  & 0.03   & $ 0.129$ & $-0.401$ & $ 0.291$ & $0.003$ & $ 0.559 \sim 0.987$ \\
$J - K$  & $K_s$$-$$K$  & 0.0001 & $-0.010$ & $ 0.042$ & $-0.019$ & $0.001$ & $ 0.332 \sim 0.772$ \\
$J - K$  & $K_s$$-$$K$  & 0.001  & $ 0.018$ & $-0.059$ & $ 0.055$ & $0.002$ & $ 0.338 \sim 0.893$ \\
$J - K$  & $K_s$$-$$K$  & 0.019  & $ 0.055$ & $-0.182$ & $ 0.134$ & $0.003$ & $ 0.539 \sim 0.992$ \\
$J - K$  & $K_s$$-$$K$  & 0.03   & $ 0.073$ & $-0.229$ & $ 0.160$ & $0.002$ & $ 0.559 \sim 0.987$ \\
$J - K$  & F205W$-$$K$  & 0.0001 & $-0.049$ & $ 0.082$ & $ 0.000$ & $0.001$ & $ 0.332 \sim 0.772$ \\
$J - K$  & F205W$-$$K$  & 0.001  & $ 0.026$ & $-0.203$ & $ 0.217$ & $0.004$ & $ 0.338 \sim 0.893$ \\
$J - K$  & F205W$-$$K$  & 0.019  & $ 0.127$ & $-0.490$ & $ 0.386$ & $0.007$ & $ 0.539 \sim 0.992$ \\
$J - K$  & F205W$-$$K$  & 0.03   & $ 0.173$ & $-0.601$ & $ 0.447$ & $0.004$ & $ 0.559 \sim 0.987$ \\
$J - K$  & F222M$-$$K$  & 0.0001 & $-0.026$ & $-0.023$ & $ 0.003$ & $0.001$ & $ 0.332 \sim 0.772$ \\
$J - K$  & F222M$-$$K$  & 0.001  & $-0.027$ & $-0.016$ & $-0.007$ & $0.001$ & $ 0.338 \sim 0.893$ \\
$J - K$  & F222M$-$$K$  & 0.019  & $-0.030$ & $-0.027$ & $ 0.002$ & $0.001$ & $ 0.539 \sim 0.992$ \\
$J - K$  & F222M$-$$K$  & 0.03   & $-0.035$ & $-0.020$ & $-0.002$ & $0.001$ & $ 0.559 \sim 0.987$ \\
$J - K'$  & $K$ $-$$K'$  & 0.0001 & $ 0.012$ & $-0.050$ & $ 0.004$ & $0.001$ & $ 0.328 \sim 0.750$ \\
$J - K'$  & $K$ $-$$K'$  & 0.001  & $-0.037$ & $ 0.138$ & $-0.140$ & $0.003$ & $ 0.334 \sim 0.871$ \\
$J - K'$  & $K$ $-$$K'$  & 0.019  & $-0.127$ & $ 0.401$ & $-0.300$ & $0.005$ & $ 0.543 \sim 0.975$ \\
$J - K'$  & $K$ $-$$K'$  & 0.03   & $-0.174$ & $ 0.520$ & $-0.369$ & $0.003$ & $ 0.593 \sim 0.971$ \\
$J - K'$  & $K_s$$-$$K'$  & 0.0001 & $ 0.002$ & $-0.006$ & $-0.017$ & $0.001$ & $ 0.328 \sim 0.750$ \\
$J - K'$  & $K_s$$-$$K'$  & 0.001  & $-0.018$ & $ 0.074$ & $-0.080$ & $0.003$ & $ 0.334 \sim 0.871$ \\
$J - K'$  & $K_s$$-$$K'$  & 0.019  & $-0.060$ & $ 0.184$ & $-0.143$ & $0.002$ & $ 0.543 \sim 0.975$ \\
$J - K'$  & $K_s$$-$$K'$  & 0.03   & $-0.080$ & $ 0.233$ & $-0.172$ & $0.001$ & $ 0.593 \sim 0.971$ \\
$J - K'$  & F205W$-$$K'$  & 0.0001 & $-0.038$ & $ 0.036$ & $ 0.004$ & $0.001$ & $ 0.328 \sim 0.750$ \\
$J - K'$  & F205W$-$$K'$  & 0.001  & $-0.006$ & $-0.087$ & $ 0.101$ & $0.002$ & $ 0.334 \sim 0.871$ \\
$J - K'$  & F205W$-$$K'$  & 0.019  & $ 0.040$ & $-0.202$ & $ 0.166$ & $0.003$ & $ 0.543 \sim 0.975$ \\
$J - K'$  & F205W$-$$K'$  & 0.03   & $ 0.077$ & $-0.286$ & $ 0.211$ & $0.002$ & $ 0.593 \sim 0.971$ \\
$J - K'$  & F222M$-$$K'$  & 0.0001 & $-0.013$ & $-0.074$ & $ 0.007$ & $0.001$ & $ 0.328 \sim 0.750$ \\
$J - K'$  & F222M$-$$K'$  & 0.001  & $-0.065$ & $ 0.125$ & $-0.150$ & $0.004$ & $ 0.334 \sim 0.871$ \\
$J - K'$  & F222M$-$$K'$  & 0.019  & $-0.160$ & $ 0.383$ & $-0.305$ & $0.004$ & $ 0.543 \sim 0.975$ \\
$J - K'$  & F222M$-$$K'$  & 0.03   & $-0.214$ & $ 0.513$ & $-0.380$ & $0.003$ & $ 0.593 \sim 0.971$ \\
$J - K_s$ & $K$ $-$$K_s$ & 0.0001 & $ 0.010$ & $-0.043$ & $ 0.020$ & $0.001$ & $ 0.330 \sim 0.761$ \\
$J - K_s$ & $K$ $-$$K_s$ & 0.001  & $-0.018$ & $ 0.060$ & $-0.056$ & $0.002$ & $ 0.335 \sim 0.885$ \\
$J - K_s$ & $K$ $-$$K_s$ & 0.019  & $-0.060$ & $ 0.197$ & $-0.143$ & $0.003$ & $ 0.544 \sim 0.989$ \\
$J - K_s$ & $K$ $-$$K_s$ & 0.03   & $-0.084$ & $ 0.257$ & $-0.177$ & $0.002$ & $ 0.594 \sim 0.986$ \\
$J - K_s$ & $K'$ $-$$K_s$ & 0.0001 & $-0.002$ & $ 0.006$ & $ 0.016$ & $0.001$ & $ 0.330 \sim 0.761$ \\
$J - K_s$ & $K'$ $-$$K_s$ & 0.001  & $ 0.017$ & $-0.069$ & $ 0.075$ & $0.003$ & $ 0.335 \sim 0.885$ \\
$J - K_s$ & $K'$ $-$$K_s$ & 0.019  & $ 0.054$ & $-0.165$ & $ 0.128$ & $0.002$ & $ 0.544 \sim 0.989$ \\
$J - K_s$ & $K'$ $-$$K_s$ & 0.03   & $ 0.070$ & $-0.204$ & $ 0.151$ & $0.001$ & $ 0.594 \sim 0.986$ \\
$J - K_s$ & F205W$-$$K_s$ & 0.0001 & $-0.040$ & $ 0.042$ & $ 0.019$ & $0.001$ & $ 0.330 \sim 0.761$ \\
$J - K_s$ & F205W$-$$K_s$ & 0.001  & $ 0.010$ & $-0.149$ & $ 0.168$ & $0.003$ & $ 0.335 \sim 0.885$ \\
$J - K_s$ & F205W$-$$K_s$ & 0.019  & $ 0.085$ & $-0.343$ & $ 0.276$ & $0.005$ & $ 0.544 \sim 0.989$ \\
$J - K_s$ & F205W$-$$K_s$ & 0.03   & $ 0.134$ & $-0.454$ & $ 0.337$ & $0.003$ & $ 0.594 \sim 0.986$ \\
$J - K_s$ & F222M$-$$K_s$ & 0.0001 & $-0.015$ & $-0.068$ & $ 0.024$ & $0.001$ & $ 0.330 \sim 0.761$ \\
$J - K_s$ & F222M$-$$K_s$ & 0.001  & $-0.045$ & $ 0.046$ & $-0.064$ & $0.002$ & $ 0.335 \sim 0.885$ \\
$J - K_s$ & F222M$-$$K_s$ & 0.019  & $-0.092$ & $ 0.174$ & $-0.144$ & $0.003$ & $ 0.544 \sim 0.989$ \\
$J - K_s$ & F222M$-$$K_s$ & 0.03   & $-0.122$ & $ 0.245$ & $-0.184$ & $0.002$ & $ 0.594 \sim 0.986$ \\
F110W$-$F205W & $K$ $-$F205W & 0.0001 & $ 0.054$ & $-0.071$ & $-0.000$ & $0.001$ & $ 0.454 \sim 0.961$ \\
F110W$-$F205W & $K$ $-$F205W & 0.001  & $-0.050$ & $ 0.223$ & $-0.176$ & $0.005$ & $ 0.484 \sim 1.107$ \\
F110W$-$F205W & $K$ $-$F205W & 0.019  & $-0.162$ & $ 0.460$ & $-0.278$ & $0.008$ & $ 0.724 \sim 1.252$ \\
F110W$-$F205W & $K$ $-$F205W & 0.03   & $-0.231$ & $ 0.584$ & $-0.328$ & $0.005$ & $ 0.784 \sim 1.257$ \\
F110W$-$F205W & $K'$ $-$F205W & 0.0001 & $ 0.040$ & $-0.030$ & $-0.003$ & $0.001$ & $ 0.454 \sim 0.961$ \\
F110W$-$F205W & $K'$ $-$F205W & 0.001  & $-0.002$ & $ 0.088$ & $-0.074$ & $0.002$ & $ 0.484 \sim 1.107$ \\
F110W$-$F205W & $K'$ $-$F205W & 0.019  & $-0.036$ & $ 0.150$ & $-0.097$ & $0.003$ & $ 0.724 \sim 1.252$ \\
F110W$-$F205W & $K'$ $-$F205W & 0.03   & $-0.068$ & $ 0.203$ & $-0.118$ & $0.002$ & $ 0.784 \sim 1.257$ \\
F110W$-$F205W & $K_s$$-$F205W & 0.0001 & $ 0.042$ & $-0.033$ & $-0.015$ & $0.001$ & $ 0.454 \sim 0.961$ \\
F110W$-$F205W & $K_s$$-$F205W & 0.001  & $-0.026$ & $ 0.162$ & $-0.133$ & $0.004$ & $ 0.484 \sim 1.107$ \\
F110W$-$F205W & $K_s$$-$F205W & 0.019  & $-0.095$ & $ 0.291$ & $-0.182$ & $0.005$ & $ 0.724 \sim 1.252$ \\
F110W$-$F205W & $K_s$$-$F205W & 0.03   & $-0.141$ & $ 0.370$ & $-0.214$ & $0.003$ & $ 0.784 \sim 1.257$ \\
F110W$-$F205W & F222M$-$F205W & 0.0001 & $ 0.030$ & $-0.091$ & $ 0.002$ & $0.001$ & $ 0.454 \sim 0.961$ \\
F110W$-$F205W & F222M$-$F205W & 0.001  & $-0.079$ & $ 0.218$ & $-0.186$ & $0.005$ & $ 0.484 \sim 1.107$ \\
F110W$-$F205W & F222M$-$F205W & 0.019  & $-0.190$ & $ 0.438$ & $-0.277$ & $0.008$ & $ 0.724 \sim 1.252$ \\
F110W$-$F205W & F222M$-$F205W & 0.03   & $-0.266$ & $ 0.569$ & $-0.330$ & $0.005$ & $ 0.784 \sim 1.257$ \\
F110W$-$F222M & $K$ $-$F222M & 0.0001 & $ 0.025$ & $ 0.018$ & $-0.002$ & $0.001$ & $ 0.465 \sim 1.017$ \\
F110W$-$F222M & $K$ $-$F222M & 0.001  & $ 0.027$ & $ 0.010$ & $ 0.006$ & $0.001$ & $ 0.496 \sim 1.170$ \\
F110W$-$F222M & $K$ $-$F222M & 0.019  & $ 0.025$ & $ 0.029$ & $-0.005$ & $0.001$ & $ 0.742 \sim 1.322$ \\
F110W$-$F222M & $K$ $-$F222M & 0.03   & $ 0.031$ & $ 0.023$ & $-0.002$ & $0.001$ & $ 0.806 \sim 1.327$ \\
F110W$-$F222M & $K'$ $-$F222M & 0.0001 & $ 0.011$ & $ 0.055$ & $-0.003$ & $0.001$ & $ 0.465 \sim 1.017$ \\
F110W$-$F222M & $K'$ $-$F222M & 0.001  & $ 0.069$ & $-0.103$ & $ 0.089$ & $0.004$ & $ 0.496 \sim 1.170$ \\
F110W$-$F222M & $K'$ $-$F222M & 0.019  & $ 0.122$ & $-0.208$ & $ 0.131$ & $0.004$ & $ 0.742 \sim 1.322$ \\
F110W$-$F222M & $K'$ $-$F222M & 0.03   & $ 0.152$ & $-0.259$ & $ 0.151$ & $0.003$ & $ 0.806 \sim 1.327$ \\
F110W$-$F222M & $K_s$$-$F222M & 0.0001 & $ 0.013$ & $ 0.053$ & $-0.014$ & $0.001$ & $ 0.465 \sim 1.017$ \\
F110W$-$F222M & $K_s$$-$F222M & 0.001  & $ 0.048$ & $-0.043$ & $ 0.042$ & $0.002$ & $ 0.496 \sim 1.170$ \\
F110W$-$F222M & $K_s$$-$F222M & 0.019  & $ 0.078$ & $-0.103$ & $ 0.068$ & $0.003$ & $ 0.742 \sim 1.322$ \\
F110W$-$F222M & $K_s$$-$F222M & 0.03   & $ 0.100$ & $-0.140$ & $ 0.082$ & $0.002$ & $ 0.806 \sim 1.327$ \\
F110W$-$F222M & F205W$-$F222M & 0.0001 & $-0.028$ & $ 0.083$ & $-0.001$ & $0.001$ & $ 0.465 \sim 1.017$ \\
F110W$-$F222M & F205W$-$F222M & 0.001  & $ 0.065$ & $-0.173$ & $ 0.148$ & $0.005$ & $ 0.496 \sim 1.170$ \\
F110W$-$F222M & F205W$-$F222M & 0.019  & $ 0.139$ & $-0.313$ & $ 0.200$ & $0.006$ & $ 0.742 \sim 1.322$ \\
F110W$-$F222M & F205W$-$F222M & 0.03   & $ 0.195$ & $-0.404$ & $ 0.235$ & $0.004$ & $ 0.806 \sim 1.327$ \\
\enddata
\tablecomments{Magnitude differences are fitted to a function
${\rm [Mag\, Diff]} = c_0 + c_1[{\rm Color}] + c_2[{\rm Color}]^2$.
Only the data points that have $\log T_{eff} \ge
3500$~K and $\log g \ge 0$ were considered for the fitting.}
\tablenotetext{a}{The largest absolute residual.}
\tablenotetext{b}{Color range where the fit is valid.}
\end{deluxetable}

\clearpage
\begin{deluxetable}{ccccccc}
\tablecolumns{7}
\tablewidth{0pt}
\tablecaption{
\label{table:lambdac}Central Wavelength $\lambda_c$ ($\mu$m)}
\tablehead{
\colhead{$J$} &
\colhead{$K$} &
\colhead{$K'$} &
\colhead{$K_s$} &
\colhead{F110W} &
\colhead{F205W} &
\colhead{F222M}
}
\startdata
1.237 &  2.212 &  2.114 &  2.160 &  1.140 &  2.079 &  2.219 \\
\enddata
\tablecomments{$\lambda_c$ is defined by eq. (8) of Paper I.}
\end{deluxetable}

\clearpage
\begin{deluxetable}{lcccccc}
\tabletypesize{\scriptsize}
\tablecolumns{7}
\tablewidth{0pt}
\tablecaption{
\label{table:alpha_eff}Averages and Standard Deviations of $\alpha_{eff}$
Values}
\tablehead{
\multicolumn{2}{c}{Isochrone Model} &
\colhead{} &
\colhead{} &
\colhead{} &
\colhead{} &
\colhead{} \\ \cline{1-2}
\colhead{$Z$} &
\colhead{Age} &
\colhead{$J-K$} &
\colhead{$J-K'$} &
\colhead{$J-K_s$} &
\colhead{F110W$-$F205W} &
\colhead{F110W$-$F222M}
}
\startdata
  0.0001 &             $10^7$ &  1.610$\pm$0.000 &  1.608$\pm$0.000 &  1.610$\pm$0.000 &  1.479$\pm$0.001 &  1.500$\pm$0.002 \\
  0.0001 &             $10^8$ &  1.608$\pm$0.003 &  1.605$\pm$0.004 &  1.607$\pm$0.004 &  1.467$\pm$0.011 &  1.486$\pm$0.011 \\
  0.0001 &             $10^9$ &  1.600$\pm$0.004 &  1.597$\pm$0.005 &  1.598$\pm$0.004 &  1.440$\pm$0.013 &  1.460$\pm$0.012 \\
  0.0001 &          $10^{10}$ &  1.596$\pm$0.003 &  1.593$\pm$0.003 &  1.595$\pm$0.003 &  1.429$\pm$0.008 &  1.450$\pm$0.007 \\
  0.001  &  $6.3 \times 10^7$ &  1.608$\pm$0.004 &  1.605$\pm$0.004 &  1.607$\pm$0.004 &  1.467$\pm$0.012 &  1.487$\pm$0.012 \\
  0.001  &             $10^8$ &  1.605$\pm$0.006 &  1.603$\pm$0.006 &  1.604$\pm$0.006 &  1.459$\pm$0.017 &  1.478$\pm$0.017 \\
  0.001  &             $10^9$ &  1.595$\pm$0.004 &  1.593$\pm$0.003 &  1.595$\pm$0.003 &  1.430$\pm$0.008 &  1.451$\pm$0.007 \\
  0.001  &          $10^{10}$ &  1.591$\pm$0.005 &  1.590$\pm$0.003 &  1.592$\pm$0.003 &  1.421$\pm$0.010 &  1.444$\pm$0.008 \\
  0.019  &             $10^7$ &  1.610$\pm$0.000 &  1.608$\pm$0.000 &  1.610$\pm$0.000 &  1.478$\pm$0.001 &  1.498$\pm$0.002 \\
  0.019  &             $10^8$ &  1.599$\pm$0.011 &  1.599$\pm$0.009 &  1.600$\pm$0.009 &  1.448$\pm$0.028 &  1.470$\pm$0.025 \\
  0.019  &             $10^9$ &  1.588$\pm$0.006 &  1.590$\pm$0.004 &  1.590$\pm$0.004 &  1.418$\pm$0.012 &  1.442$\pm$0.010 \\
  0.019  &          $10^{10}$ &  1.582$\pm$0.005 &  1.586$\pm$0.003 &  1.587$\pm$0.003 &  1.406$\pm$0.011 &  1.432$\pm$0.009 \\
  0.03   &  $6.3 \times 10^7$ &  1.606$\pm$0.007 &  1.605$\pm$0.006 &  1.606$\pm$0.006 &  1.465$\pm$0.019 &  1.485$\pm$0.018 \\
  0.03   &             $10^8$ &  1.599$\pm$0.012 &  1.599$\pm$0.009 &  1.600$\pm$0.010 &  1.448$\pm$0.029 &  1.470$\pm$0.026 \\
  0.03   &             $10^9$ &  1.586$\pm$0.006 &  1.589$\pm$0.004 &  1.590$\pm$0.004 &  1.415$\pm$0.012 &  1.440$\pm$0.010 \\
  0.03   &          $10^{10}$ &  1.581$\pm$0.005 &  1.585$\pm$0.003 &  1.586$\pm$0.003 &  1.403$\pm$0.011 &  1.429$\pm$0.009 \\
\enddata
\tablecomments{Data are presented in the form of average $\pm$ standard
deviation.  The average and standard deviation values are calculated from
the data points of each isochrone whose intrinsic $K$ magnitudes are
between $-6$ and 0~mag.}
\end{deluxetable}

\clearpage
\begin{deluxetable}{lccrrrrrcrrrrr}
\tabletypesize{\tiny}
\tablecolumns{14}
\tablewidth{0pt}
\tablecaption{\label{table:nonlin}Extinction Behavior of {\it HST} NICMOS
Filter Pairs}
\tablehead{
\multicolumn{2}{c}{Isochrone Model} &
\colhead{} &
\multicolumn{5}{c}{F110W$-$F205W} &
\colhead{} &
\multicolumn{5}{c}{F110W$-$F222M} \\ \cline{1-2} \cline{4-8} \cline{10-14}
\colhead{$Z$} &
\colhead{Age} &
\colhead{} &
\colhead{$c_0$} &
\colhead{$c_1$} &
\colhead{$c_2$} &
\colhead{$c_3$} &
\colhead{$\sigma(A^{est})$} &
\colhead{} &
\colhead{$c_0$} &
\colhead{$c_1$} &
\colhead{$c_2$} &
\colhead{$c_3$} &
\colhead{$\sigma(A^{est})$}
}
\startdata
  0.0001  &            $10^7$ & &    7.07E-04 &    2.07E-01 & $-$1.08E-01 &    7.77E-03 &  0.011 & & $-$9.83E-05 &    2.67E-01 & $-$1.18E-01 &    7.01E-03 &  0.013 \nl
  0.0001  &            $10^8$ & &    9.78E-04 &    1.75E-01 & $-$1.05E-01 &    7.70E-03 &  0.077 & &    1.18E-04 &    2.37E-01 & $-$1.16E-01 &    7.07E-03 &  0.080 \nl
  0.0001  &            $10^9$ & &    1.61E-03 &    1.14E-01 & $-$1.00E-01 &    7.88E-03 &  0.082 & &    6.21E-04 &    1.84E-01 & $-$1.14E-01 &    7.41E-03 &  0.083 \nl
  0.0001  &         $10^{10}$ & &    1.82E-03 &    8.93E-02 & $-$9.82E-02 &    7.95E-03 &  0.049 & &    7.90E-04 &    1.62E-01 & $-$1.13E-01 &    7.53E-03 &  0.050 \nl
  0.001   & $6.3 \times 10^7$ & &    9.79E-04 &    1.75E-01 & $-$1.05E-01 &    7.69E-03 &  0.082 & &    1.01E-04 &    2.38E-01 & $-$1.16E-01 &    7.05E-03 &  0.084 \nl
  0.001   &            $10^8$ & &    1.16E-03 &    1.58E-01 & $-$1.04E-01 &    7.77E-03 &  0.115 & &    2.30E-04 &    2.21E-01 & $-$1.16E-01 &    7.17E-03 &  0.118 \nl
  0.001   &            $10^9$ & &    1.75E-03 &    9.25E-02 & $-$9.86E-02 &    7.93E-03 &  0.049 & &    6.72E-04 &    1.66E-01 & $-$1.13E-01 &    7.53E-03 &  0.047 \nl
  0.001   &         $10^{10}$ & &    1.86E-03 &    7.52E-02 & $-$9.74E-02 &    7.98E-03 &  0.057 & &    8.48E-04 &    1.51E-01 & $-$1.12E-01 &    7.57E-03 &  0.054 \nl
  0.019   &            $10^7$ & &    7.77E-04 &    2.03E-01 & $-$1.08E-01 &    7.74E-03 &  0.010 & & $-$7.16E-05 &    2.64E-01 & $-$1.18E-01 &    7.01E-03 &  0.011 \nl
  0.019   &            $10^8$ & &    1.33E-03 &    1.36E-01 & $-$1.02E-01 &    7.82E-03 &  0.172 & &    3.57E-04 &    2.05E-01 & $-$1.15E-01 &    7.23E-03 &  0.171 \nl
  0.019   &            $10^9$ & &    1.93E-03 &    6.95E-02 & $-$9.71E-02 &    7.97E-03 &  0.068 & &    8.10E-04 &    1.46E-01 & $-$1.12E-01 &    7.55E-03 &  0.066 \nl
  0.019   &         $10^{10}$ & &    2.07E-03 &    4.13E-02 & $-$9.42E-02 &    7.93E-03 &  0.064 & &    8.75E-04 &    1.22E-01 & $-$1.09E-01 &    7.48E-03 &  0.062 \nl
  0.03    & $6.3 \times 10^7$ & &    1.03E-03 &    1.72E-01 & $-$1.05E-01 &    7.73E-03 &  0.120 & &    1.43E-04 &    2.35E-01 & $-$1.16E-01 &    7.06E-03 &  0.122 \nl
  0.03    &            $10^8$ & &    1.30E-03 &    1.36E-01 & $-$1.02E-01 &    7.83E-03 &  0.182 & &    3.61E-04 &    2.06E-01 & $-$1.15E-01 &    7.25E-03 &  0.180 \nl
  0.03    &            $10^9$ & &    1.92E-03 &    6.25E-02 & $-$9.61E-02 &    7.93E-03 &  0.071 & &    8.03E-04 &    1.40E-01 & $-$1.11E-01 &    7.51E-03 &  0.068 \nl
  0.03    &         $10^{10}$ & &    2.04E-03 &    3.35E-02 & $-$9.28E-02 &    7.84E-03 &  0.063 & &    8.45E-04 &    1.15E-01 & $-$1.08E-01 &    7.44E-03 &  0.063 \nl
\enddata
\tablecomments{Coefficients of best-fit third-order polynomials for the
extinction curves in Figure~\ref{fig:adiff1} for {\it HST} NICMOS filter pairs.
The difference of the estimated extinction and the true extinction is fitted
to a function $[A_Y^{est}-A_Y] = c_0 + c_1[A_Y^{est}] + c_2[A_Y^{est}]^2 +
c_3[A_Y^{est}]^3$;  $\sigma$($A^{est}$) is the average of the standard
deviations of $A_Y^{est}-A_Y$ values.}
\end{deluxetable}

\clearpage

\clearpage

\end{document}